\documentclass[prd,amsmath,aps,floats,amssymb, floatfix, superscriptaddress, nofootinbib]{revtex4-1}

\usepackage{graphicx}
\usepackage{wasysym}
\usepackage{amsfonts}
\usepackage{subfigure}
\usepackage{hyperref}
\usepackage{color}        
\usepackage{ulem}
\usepackage{tabu}
\usepackage{tabularx}
\usepackage{tabulary}
\usepackage{widetable}
\usepackage{multirow}
\usepackage[utf8]{inputenc}

\newcommand{\mr}{\mathrm}
\newcommand{\ensav}[1]{\left\langle #1 \right\rangle}

\def\be{\begin{equation}}
\def\ee{\end{equation}}
\def\bea{\begin{eqnarray}}
\def\eea{\end{eqnarray}}
\def\vs{\nonumber\\}
\def\aap{A\&A}
\def\apj{ApJ}

\def\mnras{MNRAS}
\def\aj{AJ}

\def\physrep{Phys.~Rep.}

\begin{document}

\title{Joint Analysis of Galaxy-Galaxy Lensing and Galaxy Clustering:\\ Methodology and Forecasts for DES}

\author
{Y.~Park}
\email[Corresponding author:]{youngsoo@uchicago.edu}
\affiliation{Kavli Institute for Cosmological Physics, University of Chicago, Chicago, IL 60637, USA}
\affiliation{Department of Physics, University of Chicago, Chicago, IL 60637, USA}

\author
{E.~Krause}
\email[Corresponding author:]{lise@slac.stanford.edu}
\affiliation{Kavli Institute for Particle Astrophysics \& Cosmology, P. O. Box 2450, Stanford University, Stanford, CA 94305, USA}

\author
{S.~Dodelson}
\affiliation{Fermi National Accelerator Laboratory, P. O. Box 500, Batavia, IL 60510, USA}
\affiliation{Kavli Institute for Cosmological Physics, University of Chicago, Chicago, IL 60637, USA}

\author
{B.~Jain}
\affiliation{Department of Physics and Astronomy, University of Pennsylvania, Philadelphia, PA 19104, USA}

\author
{A.~Amara}
\affiliation{Department of Physics, ETH Zurich, Wolfgang-Pauli-Strasse 16, CH-8093 Zurich, Switzerland}

\author
{M.~R.~Becker}
\affiliation{Department of Physics, Stanford University, 382 Via Pueblo Mall, Stanford, CA 94305, USA}
\affiliation{Kavli Institute for Particle Astrophysics \& Cosmology, P. O. Box 2450, Stanford University, Stanford, CA 94305, USA}

\author
{S.~L.~Bridle}
\affiliation{Jodrell Bank Center for Astrophysics, School of Physics and Astronomy, University of Manchester, Oxford Road, Manchester, M13 9PL, UK}

\author
{J.~Clampitt}
\affiliation{Department of Physics and Astronomy, University of Pennsylvania, Philadelphia, PA 19104, USA}

\author
{M.~Crocce}
\affiliation{Institut de Ci\`encies de l'Espai, IEEC-CSIC, Campus UAB, Carrer de Can Magrans, s/n,  08193 Bellaterra, Barcelona, Spain}

\author
{P.~Fosalba}
\affiliation{Institut de Ci\`encies de l'Espai, IEEC-CSIC, Campus UAB, Carrer de Can Magrans, s/n,  08193 Bellaterra, Barcelona, Spain}

\author
{E.~Gaztanaga}
\affiliation{Institut de Ci\`encies de l'Espai, IEEC-CSIC, Campus UAB, Carrer de Can Magrans, s/n,  08193 Bellaterra, Barcelona, Spain}

\author
{K.~Honscheid}
\affiliation{Center for Cosmology and Astro-Particle Physics, The Ohio State University, Columbus, OH 43210, USA}
\affiliation{Department of Physics, The Ohio State University, Columbus, OH 43210, USA}

\author
{E.~Rozo}
\affiliation{Department of Physics, University of Arizona, Tucson, AZ 85721, USA}

\author
{F.~Sobreira}
\affiliation{Fermi National Accelerator Laboratory, P. O. Box 500, Batavia, IL 60510, USA}
\affiliation{
Laborat\'orio Interinstitucional de e-Astronomia - LIneA, Rua Gal.
Jos\'e Cristino 77, Rio de Janeiro, RJ - 20921-400, Brazil
}

\author
{C.~S\'{a}nchez}
\affiliation{Institut de F\'{\i}sica d'Altes Energies, Universitat Aut\`onoma de Barcelona, E-08193 Bellaterra, Barcelona, Spain}

\author
{R.~H.~Wechsler}
\affiliation{Department of Physics, Stanford University, 382 Via Pueblo Mall, Stanford, CA 94305, USA}
\affiliation{Kavli Institute for Particle Astrophysics \& Cosmology, P. O. Box 2450, Stanford University, Stanford, CA 94305, USA}
\affiliation{SLAC National Accelerator Laboratory, Menlo Park, CA 94025, USA}

\author
{T.~Abbott}
\affiliation{Cerro Tololo Inter-American Observatory, National Optical Astronomy Observatory, Casilla 603, La Serena, Chile}

\author
{F.~B.~Abdalla}
\affiliation{Department of Physics \& Astronomy, University College London, Gower Street, London, WC1E 6BT, UK}

\author
{S.~Allam}
\affiliation{Fermi National Accelerator Laboratory, P. O. Box 500, Batavia, IL 60510, USA}

\author
{A.~Benoit-L{\'e}vy}
\affiliation{Department of Physics \& Astronomy, University College London, Gower Street, London, WC1E 6BT, UK}

\author
{E.~Bertin}
\affiliation{CNRS, UMR 7095, Institut d'Astrophysique de Paris, F-75014, Paris, France}
\affiliation{Sorbonne Universit\'es, UPMC Univ Paris 06, UMR 7095, Institut d'Astrophysique de Paris, F-75014, Paris, France}

\author
{D.~Brooks}
\affiliation{Department of Physics \& Astronomy, University College London, Gower Street, London, WC1E 6BT, UK}

\author
{E.~Buckley-Geer}
\affiliation{Fermi National Accelerator Laboratory, P. O. Box 500, Batavia, IL 60510, USA}

\author
{D.~L.~Burke}
\affiliation{Kavli Institute for Particle Astrophysics \& Cosmology, P. O. Box 2450, Stanford University, Stanford, CA 94305, USA}
\affiliation{SLAC National Accelerator Laboratory, Menlo Park, CA 94025, USA}

\author
{A.~Carnero~Rosell}
\affiliation{Laborat\'orio Interinstitucional de e-Astronomia - LIneA, Rua Gal. Jos\'e Cristino 77, Rio de Janeiro, RJ - 20921-400, Brazil}
\affiliation{Observat\'orio Nacional, Rua Gal. Jos\'e Cristino 77, Rio de Janeiro, RJ - 20921-400, Brazil}

\author
{M.~Carrasco~Kind}
\affiliation{Department of Astronomy, University of Illinois, 1002 W. Green Street, Urbana, IL 61801, USA}
\affiliation{National Center for Supercomputing Applications, 1205 West Clark St., Urbana, IL 61801, USA}

\author
{J.~Carretero}
\affiliation{Institut de Ci\`encies de l'Espai, IEEC-CSIC, Campus UAB, Carrer de Can Magrans, s/n,  08193 Bellaterra, Barcelona, Spain}
\affiliation{Institut de F\'{\i}sica d'Altes Energies, Universitat Aut\`onoma de Barcelona, E-08193 Bellaterra, Barcelona, Spain}

\author
{F.~J.~Castander}
\affiliation{Institut de Ci\`encies de l'Espai, IEEC-CSIC, Campus UAB, Carrer de Can Magrans, s/n,  08193 Bellaterra, Barcelona, Spain}

\author
{L.~N.~da Costa}
\affiliation{Laborat\'orio Interinstitucional de e-Astronomia - LIneA, Rua Gal. Jos\'e Cristino 77, Rio de Janeiro, RJ - 20921-400, Brazil}
\affiliation{Observat\'orio Nacional, Rua Gal. Jos\'e Cristino 77, Rio de Janeiro, RJ - 20921-400, Brazil}

\author
{D.~L.~DePoy}
\affiliation{George P. and Cynthia Woods Mitchell Institute for Fundamental Physics and Astronomy, and Department of Physics and Astronomy, Texas A\&M University, College Station, TX 77843,  USA}

\author
{S.~Desai}
\affiliation{Department of Physics, Ludwig-Maximilians-Universit\"at, Scheinerstr. 1, 81679 M\"unchen, Germany}
\affiliation{Excellence Cluster Universe, Boltzmannstr.\ 2, 85748 Garching, Germany}

\author
{J.~P.~Dietrich}
\affiliation{Excellence Cluster Universe, Boltzmannstr.\ 2, 85748 Garching, Germany}
\affiliation{Universit\"ats-Sternwarte, Fakult\"at f\"ur Physik, Ludwig-Maximilians Universit\"at M\"unchen, Scheinerstr. 1, 81679 M\"unchen, Germany}

\author
{P.~Doel}
\affiliation{Department of Physics \& Astronomy, University College London, Gower Street, London, WC1E 6BT, UK}

\author
{T.~F.~Eifler}
\affiliation{Department of Physics and Astronomy, University of Pennsylvania, Philadelphia, PA 19104, USA}
\affiliation{Jet Propulsion Laboratory, California Institute of Technology, 4800 Oak Grove Dr., Pasadena, CA 91109, USA}

\author
{A.~Fausti Neto}
\affiliation{Laborat\'orio Interinstitucional de e-Astronomia - LIneA, Rua Gal. Jos\'e Cristino 77, Rio de Janeiro, RJ - 20921-400, Brazil}

\author
{E.~Fernandez}
\affiliation{Institut de F\'{\i}sica d'Altes Energies, Universitat Aut\`onoma de Barcelona, E-08193 Bellaterra, Barcelona, Spain}

\author
{D.~A.~Finley}
\affiliation{Fermi National Accelerator Laboratory, P. O. Box 500, Batavia, IL 60510, USA}

\author
{B.~Flaugher}
\affiliation{Fermi National Accelerator Laboratory, P. O. Box 500, Batavia, IL 60510, USA}

\author
{D.~W.~Gerdes}
\affiliation{Department of Physics, University of Michigan, Ann Arbor, MI 48109, USA}

\author
{D.~Gruen}
\affiliation{Max Planck Institute for Extraterrestrial Physics, Giessenbachstrasse, 85748 Garching, Germany}
\affiliation{Universit\"ats-Sternwarte, Fakult\"at f\"ur Physik, Ludwig-Maximilians Universit\"at M\"unchen, Scheinerstr. 1, 81679 M\"unchen, Germany}

\author
{R.~A.~Gruendl}
\affiliation{Department of Astronomy, University of Illinois, 1002 W. Green Street, Urbana, IL 61801, USA}
\affiliation{National Center for Supercomputing Applications, 1205 West Clark St., Urbana, IL 61801, USA}

\author
{G.~Gutierrez}
\affiliation{Fermi National Accelerator Laboratory, P. O. Box 500, Batavia, IL 60510, USA}

\author
{D.~J.~James}
\affiliation{Cerro Tololo Inter-American Observatory, National Optical Astronomy Observatory, Casilla 603, La Serena, Chile}

\author
{S.~Kent}
\affiliation{Fermi National Accelerator Laboratory, P. O. Box 500, Batavia, IL 60510, USA}

\author
{K.~Kuehn}
\affiliation{Australian Astronomical Observatory, North Ryde, NSW 2113, Australia}

\author
{N.~Kuropatkin}
\affiliation{Fermi National Accelerator Laboratory, P. O. Box 500, Batavia, IL 60510, USA}

\author
{M.~Lima}
\affiliation{
Departamento de F\'{\i}sica Matem\'atica, 
Instituto de F\'{\i}sica, Universidade de S\~ao Paulo, 
CP 66318, CEP 05314-970, S\~ao Paulo, SP, Brazil
}
\affiliation{
Laborat\'orio Interinstitucional de e-Astronomia - LIneA, Rua Gal.
Jos\'e Cristino 77, Rio de Janeiro, RJ - 20921-400, Brazil
}

\author
{M.~A.~G.~Maia}
\affiliation{Laborat\'orio Interinstitucional de e-Astronomia - LIneA, Rua Gal. Jos\'e Cristino 77, Rio de Janeiro, RJ - 20921-400, Brazil}
\affiliation{Observat\'orio Nacional, Rua Gal. Jos\'e Cristino 77, Rio de Janeiro, RJ - 20921-400, Brazil}

\author
{J.~L.~Marshall}
\affiliation{George P. and Cynthia Woods Mitchell Institute for Fundamental Physics and Astronomy, and Department of Physics and Astronomy, Texas A\&M University, College Station, TX 77843,  USA}

\author
{P.~Melchior}
\affiliation{Center for Cosmology and Astro-Particle Physics, The Ohio State University, Columbus, OH 43210, USA}
\affiliation{Department of Physics, The Ohio State University, Columbus, OH 43210, USA}

\author
{C.~J.~Miller}
\affiliation{Department of Astronomy, University of Michigan, Ann Arbor, MI 48109, USA}
\affiliation{Department of Physics, University of Michigan, Ann Arbor, MI 48109, USA}

\author
{R.~Miquel}
\affiliation{Instituci\'o Catalana de Recerca i Estudis Avan\c{c}ats, E-08010 Barcelona, Spain}
\affiliation{Institut de F\'{\i}sica d'Altes Energies, Universitat Aut\`onoma de Barcelona, E-08193 Bellaterra, Barcelona, Spain}

\author
{R.~C.~Nichol}
\affiliation{Institute of Cosmology \& Gravitation, University of Portsmouth, Portsmouth, PO1 3FX, UK}

\author
{R.~Ogando}
\affiliation{Laborat\'orio Interinstitucional de e-Astronomia - LIneA, Rua Gal. Jos\'e Cristino 77, Rio de Janeiro, RJ - 20921-400, Brazil}
\affiliation{Observat\'orio Nacional, Rua Gal. Jos\'e Cristino 77, Rio de Janeiro, RJ - 20921-400, Brazil}

\author
{A.~A.~Plazas}
\affiliation{Jet Propulsion Laboratory, California Institute of Technology, 4800 Oak Grove Dr., Pasadena, CA 91109, USA}

\author
{N.~Roe}
\affiliation{Lawrence Berkeley National Laboratory, 1 Cyclotron Road, Berkeley, CA 94720, USA}

\author
{A.~K.~Romer}
\affiliation{Department of Physics and Astronomy, Pevensey Building, University of Sussex, Brighton, BN1 9QH, UK}

\author
{E.~S.~Rykoff}
\affiliation{Kavli Institute for Particle Astrophysics \& Cosmology, P. O. Box 2450, Stanford University, Stanford, CA 94305, USA}
\affiliation{SLAC National Accelerator Laboratory, Menlo Park, CA 94025, USA}

\author
{E.~Sanchez}
\affiliation{Centro de Investigaciones Energ\'eticas, Medioambientales y Tecnol\'ogicas (CIEMAT), Madrid, Spain}

\author
{V.~Scarpine}
\affiliation{Fermi National Accelerator Laboratory, P. O. Box 500, Batavia, IL 60510, USA}

\author
{M.~Schubnell}
\affiliation{Department of Physics, University of Michigan, Ann Arbor, MI 48109, USA}

\author
{I.~Sevilla-Noarbe}
\affiliation{Centro de Investigaciones Energ\'eticas, Medioambientales y Tecnol\'ogicas (CIEMAT), Madrid, Spain}
\affiliation{Department of Astronomy, University of Illinois, 1002 W. Green Street, Urbana, IL 61801, USA}

\author
{M.~Soares-Santos}
\affiliation{Fermi National Accelerator Laboratory, P. O. Box 500, Batavia, IL 60510, USA}

\author
{E.~Suchyta}
\affiliation{Center for Cosmology and Astro-Particle Physics, The Ohio State University, Columbus, OH 43210, USA}
\affiliation{Department of Physics, The Ohio State University, Columbus, OH 43210, USA}

\author
{M.~E.~C.~Swanson}
\affiliation{National Center for Supercomputing Applications, 1205 West Clark St., Urbana, IL 61801, USA}

\author
{G.~Tarle}
\affiliation{Department of Physics, University of Michigan, Ann Arbor, MI 48109, USA}

\author
{J.~Thaler}
\affiliation{Department of Physics, University of Illinois, 1110 W. Green St., Urbana, IL 61801, USA}

\author
{V.~Vikram}
\affiliation{Argonne National Laboratory, 9700 South Cass Avenue, Lemont, IL 60439, USA}

\author
{A.~R.~Walker}
\affiliation{Cerro Tololo Inter-American Observatory, National Optical Astronomy Observatory, Casilla 603, La Serena, Chile}

\author
{J.~Weller}
\affiliation{Excellence Cluster Universe, Boltzmannstr.\ 2, 85748 Garching, Germany}
\affiliation{Max Planck Institute for Extraterrestrial Physics, Giessenbachstrasse, 85748 Garching, Germany}
\affiliation{Universit\"ats-Sternwarte, Fakult\"at f\"ur Physik, Ludwig-Maximilians Universit\"at M\"unchen, Scheinerstr. 1, 81679 M\"unchen, Germany}

\author
{J.~Zuntz}
\affiliation{Jodrell Bank Center for Astrophysics, School of Physics and Astronomy, University of Manchester, Oxford Road, Manchester, M13 9PL, UK}

\begin{abstract}
The joint analysis of galaxy-galaxy lensing and galaxy clustering is a promising method for inferring the growth function of large scale structure. This analysis will be carried out on data from the Dark Energy Survey (DES), with its measurements of both the distribution of galaxies and the tangential shears of background galaxies induced by these foreground lenses. We develop a practical approach to modeling the assumptions
and systematic effects affecting small scale lensing, which provides halo masses, and large scale galaxy clustering. Introducing parameters that characterize the halo occupation distribution (HOD), photometric redshift uncertainties, and shear measurement errors, we study how external priors on different subsets of these parameters affect our growth constraints. Degeneracies within the HOD model, as well as between the HOD and the growth function, are identified as the dominant source of complication, with other systematic effects sub-dominant. The impact of HOD parameters and their degeneracies necessitate the detailed joint modeling of the galaxy sample that we employ. We conclude that DES data will provide powerful constraints on the evolution of structure growth in the universe, conservatively/optimistically constraining the growth function to 7.9\%/4.8\% with its first-year data that covered over 1000 square degrees, and to 3.9\%/2.3\% with its full five-year data that will survey 5000 square degrees, including both statistical and systematic uncertainties.
\end{abstract}

\maketitle

\section{Introduction}

 Evidence from multiple probes now points to an accelerated expansion of the Universe. Distant Type Ia supernovae are fainter than they would be if the Universe were decelerating~\citep{Riess98,Perlmutter99}; patterns in the anisotropy of the Cosmic Microwave Background (CMB) have long been consistent with acceleration and now offer solid independent evidence \citep{planckover}; the scale of Baryonic Acoustic Oscillations in the late-time galaxy distributions also points to acceleration~\citep{boss}. Other measurements, while not providing stand-alone evidence, are nonetheless consistent with the notion that the deceleration predicted by Einstein's theory of General Relativity without a cosmological constant is not occurring today. For example, measurements of growth of structure using the abundance of massive clusters of galaxies~\citep{Benson13,Mantz14}, as well as weak lensing \citep{cfhtlens} have been found to be consistent with a model in which dark energy driving acceleration contributes roughly $70\%$ of the energy density of the Universe. The physical nature of the mechanism driving this accelerated expansion, however, is still to be determined. 

A major goal of the Dark Energy Survey
(DES) is to understand that mechanism by measuring the growth of large scale structure. Because different models predict distinct histories of structure growth in the late-time universe, constraints on growth history can lead to constraints on the mechanism responsible for cosmic acceleration. We expect the most precise  constraints to be obtained using combinations of several probes (e.g., see~\cite{DETF}), which increase the overall signal-to-noise and break parameter degeneracies -- both among the cosmological parameters of interest and the nuisance parameters that quantify systematic effects. The example we focus on here is the combination of measurements of galaxy-galaxy lensing and clustering of the lens galaxy sample, which has been suggested in the past few years by \citep{Yoo:2012,vandenBosch:2012}. By constraining the growth function with such a combined analysis, we not only constrain the parameters of the ``standard model'' of cosmology but also can detect possible deviations from the robust predictions of General Relativity and smooth dark energy models. 

In particular we implement the approach proposed in \citet{Yoo:2012}, which combines small-scale galaxy-galaxy lensing with large-scale clustering, on mock data sets designed to resemble that from DES. On large spatial scales, the galaxy overdensity is proportional to the overdensity in the total matter distribution, with the relation between the two over-densities captured by a single number, the linear bias parameter (e.g., see \citep{Mandelbaum:2013}) which is related to the masses of halos hosting the galaxy sample.
On small spatial scales the relation between galaxy and dark matter distribution is non-linear.  The small-scale dark matter distribution is assumed to follow that of a spherical halo with a universal mass profile, and the distribution of galaxies within a halo is commonly described by Halo Occupation Distributions (HOD) \citep{Berlind02}. HODs are used extensively to model galaxy-galaxy lensing and clustering \citep[e.g.][]{Leauthaud11,vandenBosch:2012}, and have been successfully applied in recent  joint analyses of galaxy-galaxy lensing and clustering \citep{Leauthaud12,Cacciato:2012,More:2014}. The insight of \citep{Yoo:2012} was that one could apply a step-by-step method to address these different scales and corresponding physics, starting by fitting lensing on small scales with a simple mass profile. Then, the inferred mass could be used to understand the large scale bias of the lensing galaxies, turning large-scale galaxy clustering measurements into direct probes of the underlying clustering of matter. By carrying out this two-step analysis with lensing galaxies in multiple redshift bins, one might therefore be able to measure the growth of the structure.

Here, we implement this idea anticipating a near future application to DES data. We find that the simple two-step approach needs to be tweaked, and that parameters accounting for the HOD modeling as well as for systematic effects need to be introduced into a full one-step analysis that includes both sets of measurements -- clustering and lensing -- in one data vector.
We develop a full analysis pipeline for this method and forecast its constraining power at different stages of the Dark Energy Survey. 
We employ a joint model for key systematic effects such as halo model assumptions and photometric redshift errors, allowing for both probes to contribute information on, and best constrain, the underlying assumptions and parameters under a realistic setting. In addition, to correctly account for the correlation between probes, we utilize the full joint covariance matrix of the two probes. We test and validate this pipeline in simulated data designed to closely mimic that obtained and expected from DES. This pipeline will be applied to DES data, so one of our goals here is to test the pipeline and the underlying algorithm: how can we optimize this joint analysis on actual survey data given the statistical uncertainties and likely sources of systematic error? Which systematic effects are most important to model accurately and which do not affect the final cosmological constraints? Most generally, how accurately should we expect to be able to extract information about the growth of cosmic structure?

The plan of the paper is as follows. Section II contains a description of the implementation: the halo model based formalism and the choice of our parameter set. The mock catalogs, measurements and tests are presented in Section III.  In Section IV we describe our likelihood analysis and details on model parametrization. We present and discuss our results in Sections V and VI.

\section{Modeling}
\label{sec:impl}

\subsection{Motivation}

The focus of this paper is to develop a pipeline that will extract information about the growth function from small-scale DES galaxy-galaxy lensing and large-scale DES galaxy clustering. 

A first attempt to implement the method described in Ref.~\citep{Yoo:2012} would be to:
\begin{enumerate}
\item Select a galaxy sample with a given luminosity cut, with a parametrized model for the mass-luminosity relation and redshift range.
\item Fit the halo-scale galaxy-galaxy lensing data, $\gamma_\mr{t}(\theta)$, with a halo mass profile to extract an estimate of the mean mass of the sample.
\item Determine the large-scale halo bias for that mass using fits from numerical simulations, e.g. \citep{Tinker:2010my}.
\item Measure the angular correlation function, $w(\theta)$, of the galaxy sample.
\item Using the inferred halo bias and external priors from, e.g., Planck \citep{Ade:2013zuv}, simultaneously fit the correlation function to a set of cosmological parameters including the growth function.
\end{enumerate}

Writing these steps down immediately reveals a number of problems. In order to carry out each of Steps 1-3, a distance-redshift relation is needed, which depends on cosmology. So in principle, one cannot fix this relation and then at the final step fit for cosmological parameters. Second, redshift bins will be determined using DES colors so will be subject to photometric redshift errors, and these affect the fits in  Steps 1, 3 and 5. Therefore uncertainties in photometric redshifts must be treated simultaneously. Finally, some information is needed about the mass-luminosity relation and particularly about the fraction of galaxies that are satellites instead of central galaxies. For these purposes a more sophisticated analysis is needed even at the outset.

We aim to maintain the basic idea of \citet{Yoo:2012} of combining small-scale galaxy galaxy lensing with large-scale galaxy clustering, while addressing the above issues. Our starting point then is the joint data vector that includes both $\gamma_\mr{t}(\theta)$ and $w(\theta)$ for the luminosity-threshold galaxy sample. To extract predictions for these statistics, we employ a halo model \citep{Cooray02} in combination with HOD modeling. Specifically, we define halos as spherical overdensities of $\Delta_m = \rho/\rho_m = 200$, and assume their densities follow the Navarro, Frenk, \& White (NFW) profiles \citep{NFW} with the \citet{Duffy08} mass--concentration relation. We use \citet{Tinker:2008ff,Tinker:2010my} fitting functions for the halo mass function and halo mass--bias relation, respectively. We then jointly model both $w(\theta)$ and $\gamma_\mr{t}(\theta)$ from this halo model picture, with added ingredients for systematic effects such as photometric redshift errors and multiplicative shear calibration. In addition to the parameters associated with the HOD modeling, the set of systematic effects, and cosmology, we introduce growth scaling parameters, denoted $A_i$, to freely scale the amplitude of the growth function in each redshift bin, rendering our analysis capable of both constraining the growth function and detecting potential deviations from $\mr{\Lambda CDM}$ structure growth. A key ingredient of this analysis is the full joint covariance matrix of the joint data vector. In treating the joint likelihood, the full joint covariance matrix allows for a proper accounting of the information in the joint data vector, especially with its off-diagonal blocks representing covariances between the two probes. 

\subsection{Halo Occupation Distribution}

\begin{figure}[ht]
\includegraphics[width=0.5\textwidth]{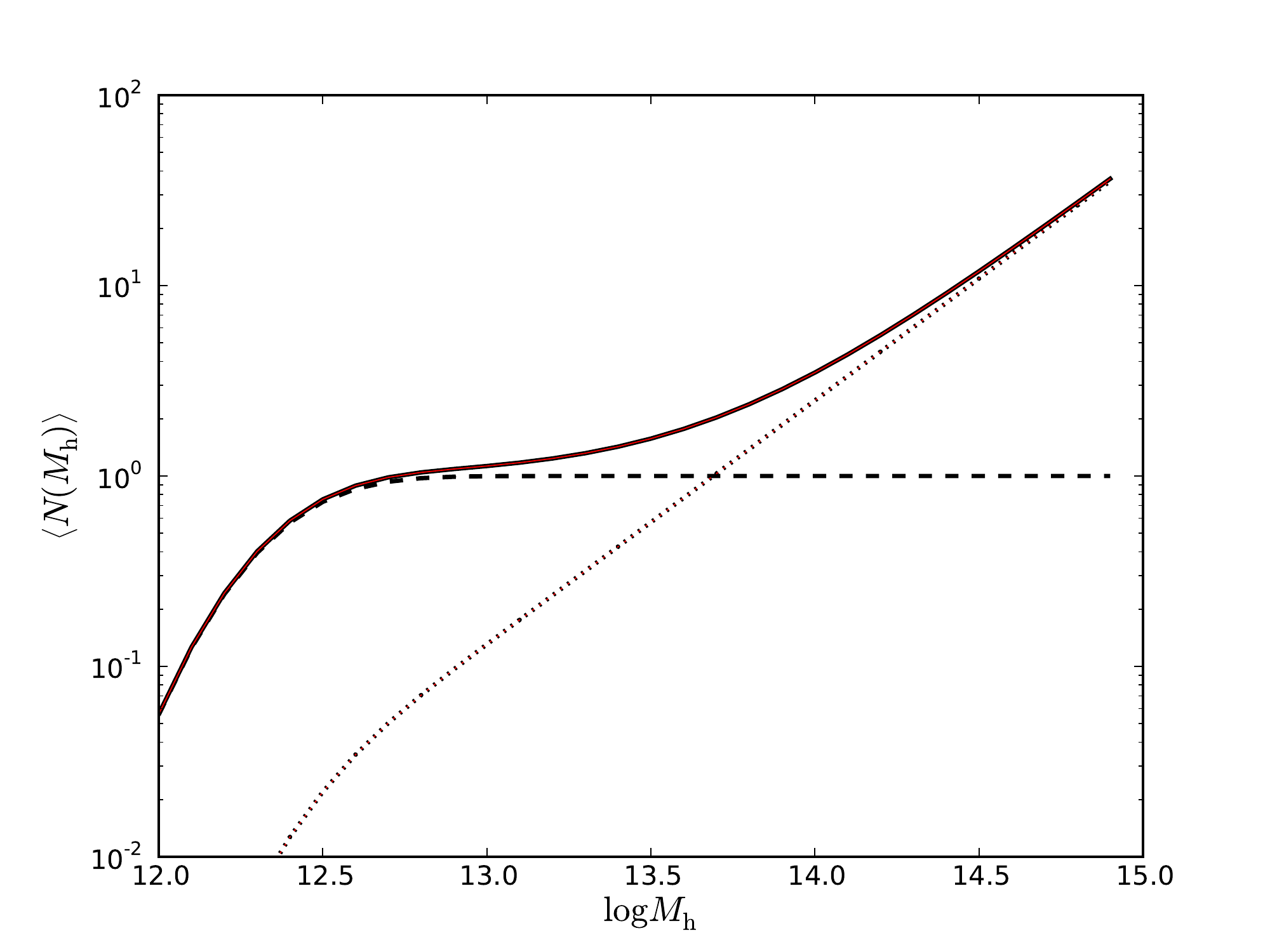}
\caption{An example of the average number of central/satellite galaxies, $\ensav{N_\mr{c/s}(M_{\mr h})}$, calculated from Eq. \ref{eq:NM2} with parameter settings $\log M_\mr{min}= 12.36$, $\log M_\mr{1}= 13.69$, $\sigma_{\log M} = 0.32$, $\alpha = 1.28$. These parameter values are selected to match our fiducial default values presented in Table \ref{tab:params}. The dashed and dotted black lines respectively represent the central and satellite galaxy counts, with the solid black line showing their sum, i.e. the total number of galaxies in a halo of mass $M_\mr h$. The solid and dashed red lines respectively represent the satellite and total counts using $\log M_0 = 8.35$ in addition, i.e. Eq. \ref{eq:NM2} before our simplification. For the galaxy sample under consideration, the effect of the satellite cut-off mass scale $M_0$ is negligible.}
\label{fig:hod}
\end{figure}

When we measure the tangential shear induced by stacked foreground halos, what is the best way to characterize our sample? Simply fitting for a single value, i.e. the mean halo mass, is not optimal because it does not fully represent the underlying mass distribution of halos, thereby leaving out information. Rather, directly modeling that underlying mass distribution by means of a halo mass function will yield a more realistic characterization of the sample. Furthermore, we observe galaxies, not halos, so in addition to the mass function we also need a recipe that connects galaxies to halos. Going from a halo mass function to a galaxy distribution requires an HOD model that describes the relation between galaxies and halo mass in terms of the probability $P(N|M_{\mr h})$ that a halo of given mass $M_{\mr h}$ contains $N$ galaxies. We separate galaxies into central and satellite galaxies. By definition, a halo contains either zero or one central galaxy, and it can only host satellite galaxies if it contains a central galaxy, which motivates the form \citep{Zheng05}
\be
 \ensav{N(M_{\mr h})} = \ensav{N_{\mr{c}}(M_{\mr h})}\left(1 + \ensav{N_{\mr{s}}(M_{\mr h})}\right),
 \label{eq:NM}
\ee
with $\ensav{N_{\mr{c/s}}(M_{\mr h})}$ the average number of central/satellite galaxies in a halo of mass $M_\mr{h}$.

For a luminosity-threshold sample (with absolute $r$-band magnitude $\mathcal M_r<\mathcal M_r^t$), the HOD for centrals and satellites is commonly parameterized as \citep[e.g.,][]{Zheng05}
\bea
\ensav{N_c(M_{\mr h}|\mathcal M_r^t)} &=& \frac{1}{2}\left[1+\mr{erf}\left(\frac{\log M_{\mr h}-\log M_{\mr{min}}}{\sigma_{\mr{log}M}}\right)\right]
\vs
\ensav{N_s(M_{\mr h}|\mathcal M_r^t)} &=& 
\left(\frac{M_{\mr h}-M_0}{M_1^{\prime}}\right)^{\alpha}\,,
\label{eq:NM2}
\eea
with model parameters $M_{\mr{min}}, M_1^{\prime}, \sigma_{\mr{log}M}, \alpha$, and all mass parameters in units of $M_\odot / h$. The central galaxy occupation function is a softened step function with transition mass scale $M_{\mr{min}}$, which is the halo mass in which the median central galaxy luminosity corresponds to the luminosity threshold, and softening parameter $\sigma_{\mr{log}M}$ which is related to the scatter between galaxy luminosity and halo mass. The normalization of the satellite occupation function, $M_1^{\prime}$, and cut-off scale $M_0$ are related to $M_1$, the mass scale at which a halo hosts at least one satellite galaxy ($\ensav{N_{\mr s}(M_1)} =1)$); finally $\alpha$ is the high-mass-end slope of the satellite occupation function. This parametrization was found to reproduce the clustering of SDSS \citep{Zehavi11} and CFHTLS \citep{Coupon:2011mz} galaxies well over a large range of luminosity thresholds and redshifts. 
To simplify this model and reduce the number of fit parameters, we ignore the satellite cut-off scale $M_0 \equiv 0$ and use a four-parameter model for luminosity threshold samples.
Fig. \ref{fig:hod} illustrates our HOD model, exhibiting the soft low-mass threshold determined by $\log M_\mr{min}$ and $\sigma_{\log M}$, the satellite onset dictated by $\log M_\mr 1$, and the rapid increase of satellite counts at the high-mass end governed by $\alpha$.
Section~\ref{sec:obs} describes the halo model that relates the HOD to galaxy-galaxy lensing and clustering observables. Throughout we use the mean value of central and satellite occupation, and assume that satellite galaxies are Poisson distributed both in number and in position with respect to halo centers to calculate second moments.

\subsection{Photometric Redshift Uncertainties and Shear Calibration}
The redshift distribution of galaxies plays a key role in projecting the 3D information to the 2D observables $w(\theta)$ and $\gamma_{\mr t}(\theta)$, as well as in interpreting the tangential shear of a source galaxy image by a lens galaxy. In photometric surveys like DES the true redshifts of observed galaxies are not available; instead, redshift values are estimated from a galaxy's brightness in different colors, known as photometric redshifts, or photo-z's, $z_{\mr{ph}}$. Galaxies with photometric redshifts within a given range are lumped into a photometric redshift bin. To infer the true redshift distribution of this bin, we convolve the conditional probability function $p(z|z_{\mr {ph}})$ with the photometric redshift distribution $n(z_{\mr {ph}})$ to calculate the true redshift distribution of the $i$-th photometric redshift bin $n_i(z)$:
\be 
 \label{eq:photoz}
n_i(z) = \int_{z_{\mr {ph}}^{\mr {min,i}}}^{z_{\mr {ph}}^{\mr {max,i}}} dz_{\mr {ph}} p(z_{\mr {ph}}|z) n(z_{\mr {ph}})\,.
\ee
We assume a Gaussian distribution of photo-zs around a true redshift value with the redshift-dependent standard error $\sigma = \sigma_z (1+z)$ and constant offset $b_z$ \citep[e.g.,][]{Ma06},
\be 
 p(z_{\mr {ph}}|z) = \frac{1}{\sqrt{2\pi}\sigma}\exp\left[-\frac{(z-z_{\mr {ph}}-b_z)^2}{2\sigma^2}\right].
\ee
This model is a somewhat idealized picture, as in reality complex galaxy spectra give rise to complicated, non-Gaussian photo-z distributions. Here we choose this simple parametrization to manifest the most important modes of error in photo-z's, also noting that the Gaussian assumption holds well for our expected candidates for lens galaxies, namely Luminous Red Galaxies (LRG's).

In addition, we consider a multiplicative calibration of the observed tangential shear as a potential source of systematic effects. so that the true shear is related to the observed shear via
\be
\ensav{\gamma_t(\theta)}_\mr{true} = (1 + m_\gamma) \ensav{\gamma_t(\theta)}_\mr{obs}.
\ee

\subsection{Growth Function Scaling}
At the linear level, the growth of structure in the universe is described by the growth function $D(z)$, normalized to be unity at $z=0$. For example, in terms of $D(z)$, the matter power spectrum $P(k,z)$ is 
\be 
 P(k,z) = D^2(z) P(k,0),
 \ee
which then enters various structure-related quantities such as the variance of matter density fluctuations on a scale $R$, $\sigma_R(z)$, and subsequently the mass function $dn/dM_\mr{h}$. For a standard flat LCDM cosmology, $D(z)$ is given by
\be 
 D^{\rm \Lambda CDM}(z) = \frac{H(z)}{H_0}\int_z^\infty \frac{dz'(1+z')}{H^3(z')}\left[\int_0^\infty \frac{dz''(1+z'')}{H^3(z'')}\right]^{-1},
 \ee
 where $H(z) = H_0 \left(\Omega_M(1+z)^3+\Omega_\Lambda\right)^{1/2}$ with the present-day Hubble constant $H_0$ and density parameters $\Omega_M$, $\Omega_\Lambda$. Therefore, these three parameters uniquely define the growth function in the LCDM scenario. Therefore, in order to capture sensitivity to possible anomalies in the growth function, we introduce free scaling parameters $A_i$ defined
 \be 
  \tilde D_i(z) = A_iD^{\rm \Lambda CDM}(z),
  \ee
 that scales the growth function for the $i$-th redshift bin in our galaxy sample. The ensuing constraints on $A_i$ capture the sensitivity of the combined probes to the amplitude of fluctuations at the redshift of interest. If the $A_i$ are fonud to differ from one at a significant level, then LCDM would be ruled out. More generally, modified gravity models make different predictions for growth than do dark energy models, so independent measures of growth such as the $A_i$ are extremely valuable ways to distinguish between these competing ideas for the cause of the cosmic acceleration.

\subsection{Observables}
\label{sec:obs}
\subsubsection{Large-Scale Galaxy Clustering}

Our analysis uses the two-point function of the galaxy distribution on scales larger than individual halos. The angular power spectrum of galaxies in a given redshift bin $i$ then depends on the linear matter power spectrum via
\be
C^i_{gg}(l) = \int dz H(z) \chi^{-2}(z)  W_{g,i}^2(z) P(k=l/\chi,z),
\ee
where $\chi(z)$ is the comoving distance out to redshift $z$, and the galaxy window function $W_{g,i}(z)$ in bin $i$ is 
\be 
W_{g,i}(z) = \frac{n_i(z)}{\bar n_i}\bar{b}_\mr{g}(z),
\ee
with $n_i(z)$ the redshift distribution inferred from photometric estimates (see Eq. \ref{eq:photoz}), and normalization factor $\bar n_i\equiv\int dz\,n_i(z)$. The mean galaxy bias $\bar{b}_\mr{g}(z)$ is given by
\be
\bar{b}_\mr{g}(z) = \frac{1}{\bar n _M}\int_0^\infty d M_{\mr h}\,\frac{dn}{dM_{\mr h}} b_{\mr h}(M_{\mr h})\big|_{z}\ensav{N(M_{\mr h}|X)} \,
\label{eq:bg}
\ee
with $X = \{M_{\rm{min}}, M_1',\sigma_{\rm{log}M},\alpha\}$ representing the HOD parameters defined in Eq.~\ref{eq:NM2}. Here, ${dn}/{dM_{\mr h}}$ and $b_{\mr h}(M_{\mr h})$ are the halo mass function and the halo mass-bias relation from \citet{Tinker:2008ff} and \citet{Tinker:2010my}, respectively. Note that as these quantities depend on $\sigma(R,z)$, they are affected by the growth scaling parameters $A_i$.  The normalization parameter $\bar{n}_M$ is given by
\be
\bar{n}_M = \int_0^\infty d M_{\mr h}\,\frac{dn}{dM_{\mr{h}}}\ensav{N(M_{\mr h}|X)}.
\label{eq:ngal}
\ee
In the flat sky limit, our observable $w(\theta)$ is related to $C_{gg}(l)$ as
\be 
w(\theta) = \int \frac{ldl}{2\pi}C_{gg}(l)J_0(l\theta),
\ee
where $J_0(l\theta)$ is the zeroth-order Bessel function.

\subsubsection{Small-Scale Galaxy-Galaxy Lensing}

The measured tangential shear $\langle\gamma_\mr{t}^{ij}(\theta)\rangle$ of foreground galaxies in redshift bin $i$ and source galaxies in redshift bin $j$ is related to the Fourier transform of the tomographic galaxy-convergence angular power spectrum, $C_{g\kappa}^{ij}(l)$ by
\be 
\langle\gamma_t^{ij}(\theta)\rangle= \int \frac{ldl}{2\pi}C_{g\kappa}^{ij}(l)J_2(l\theta)\,,
\ee
with $J_2$ the second-order Bessel function.

The angular galaxy-convergence power spectrum is an integral over the 3D galaxy-mass power spectrum; in the small angle Limber approximation,
\be 
 C_{g\kappa}^{ij}(l) = \int dz \chi^{-2}(z) \frac{n_i(z)}{\bar n_i}W_\kappa^j (z)P_\mr{gm}(k=l/\chi,z).
 \ee
Here, the lensing window function $W_\kappa^j(z)$ for source bin $j$ is
\be 
 W_\kappa^j(z) = \frac{\bar{\rho}_m (z)}{(1+z)\Sigma^j_\mr{crit}(z)},
\ee
where the critical surface density $\Sigma^j_\mr{crit}(z)$ of source bin $j$ is given by
\be 
 \left(\Sigma_\mr{crit}^j\right)^{-1}(z) = \frac{4\pi G \chi(z)}{1+z}\left[1-\chi(z) \ensav{\frac{1}{\chi(z_s)}}\right],
\ee
with $\langle\chi^{-1}(z_s)\rangle$ the mean inverse comoving distance to the source galaxies in source bin $j$.

It remains to compute the 3D galaxy-mass spectrum, which we describe using the halo model and HOD. For this analysis we will ignore the contribution of sub-halos and model the lensing signal around satellite galaxies with mis-centered NFW halos. Since we focus on small scales, we consider only the one-halo term:
\be
P_{\mr{gm}}^{1h}(k,X) = P_{\mr{cm}}(k,X) + P_{\mr{sm}}(k,X) = \frac{1}{\bar{\rho}_{\mr{m}}\bar{n}_{M}}\int d M_{\mr h} M_{\mr h} \tilde u_{\mr h}(k, M_{\mr h})\frac{dn}{dM_{\mr h}}\left[\ensav{N_{\mr c}(M_{\mr h}|X)} + \ensav{N_{\mr s}(M_{\mr h}|X)}\tilde u_{\mr s}(k, M_{\mr h})\right]\,,
\ee
where $\tilde u_{\mr{h}}(k, M_{\mr h})$ is the Fourier transform of the halo density profile of mass $M_{\mr h}$, and $\tilde u_{\mr{s}}(k, M_{\mr h})$ the Fourier transform of the spatial distribution of satellite galaxies within the halo. Here, we assume that the distribution of satellite galaxies follows the NFW profile by letting $\tilde u_{\mr{s}}=\tilde u_{\mr{h}}$, and also that central galaxies are located at the exact halo centers, i.e. without mis-centering.

\section{Mock Data}
\label{sec:data}

\subsection{DES Data Stages}

DES is an ongoing wide field multi-color imaging survey that will cover nearly 5000 square degrees of the southern sky with a limiting $i$-band magnitude of 24 by Spring 2018. Its images come from the Dark Energy Camera \cite{decam}, a 3 square degree imager on the Blanco Telescope near La Serena, Chile. Images taken by the camera to roughly comparable depths will be obtained in $g, r, i, z, Y$ bands, which will be used to characterize the positions, redshifts, and shapes of about 300 million galaxies. 
Pre-survey science verification data was taken from December 2012 to February 2013 and processed in Fall 2013. This data set, named the SVA1 data release, covers about 150 square degrees to a limiting magnitude $m_r \sim 24$ in the $r$ band.
The first year of science observations, referred to as Y1, has been released, covering over 1000 square degrees to roughly 0.5 magnitudes shallower depth. 
The complete DES dataset, to be achieved with 5 years of full data taking, is referred to as the Y5 dataset. 

To test our modeling discussed in Section \ref{sec:impl}, we construct a number of fiducial datasets from numerical simulations. In this work, we consider two different DES data stages, namely the DES Y1 and DES Y5 stages. The full range of our pipeline is tested using simulated likelihood analyses with DES Y1-like and Y5-like survey parameters and mock covariances from the DES Blind Cosmology Challenge (BCC) simulation results.

\subsection{Mock Survey Setup}
\label{subsec:y1y5data}

We make use of the DES BCC mock galaxy catalogs developed for the DES 
collaboration (\citet{Busha15}) to construct our mock surveys. From these DES galaxy mock catalogs, we construct luminosity-thresholded lens galaxy samples over two redshift bins, namely at $0.3<z<0.4$ with $\mathcal{M}_{\mathrm r} <-21.5$ and and at $0.4<z<0.5$ with $\mathcal{M}_{\mathrm r} <-22.0$. In order to obtain a realistic galaxy-galaxy lensing signal from mock catalogs with finite mass resolution the host halo of the lens galaxy needs to be resolved. Hence the luminosity thresholds for our lens samples are chosen such that central galaxies are located in resolved halos.

\begin{figure}[ht]
\includegraphics[width=0.49\textwidth]{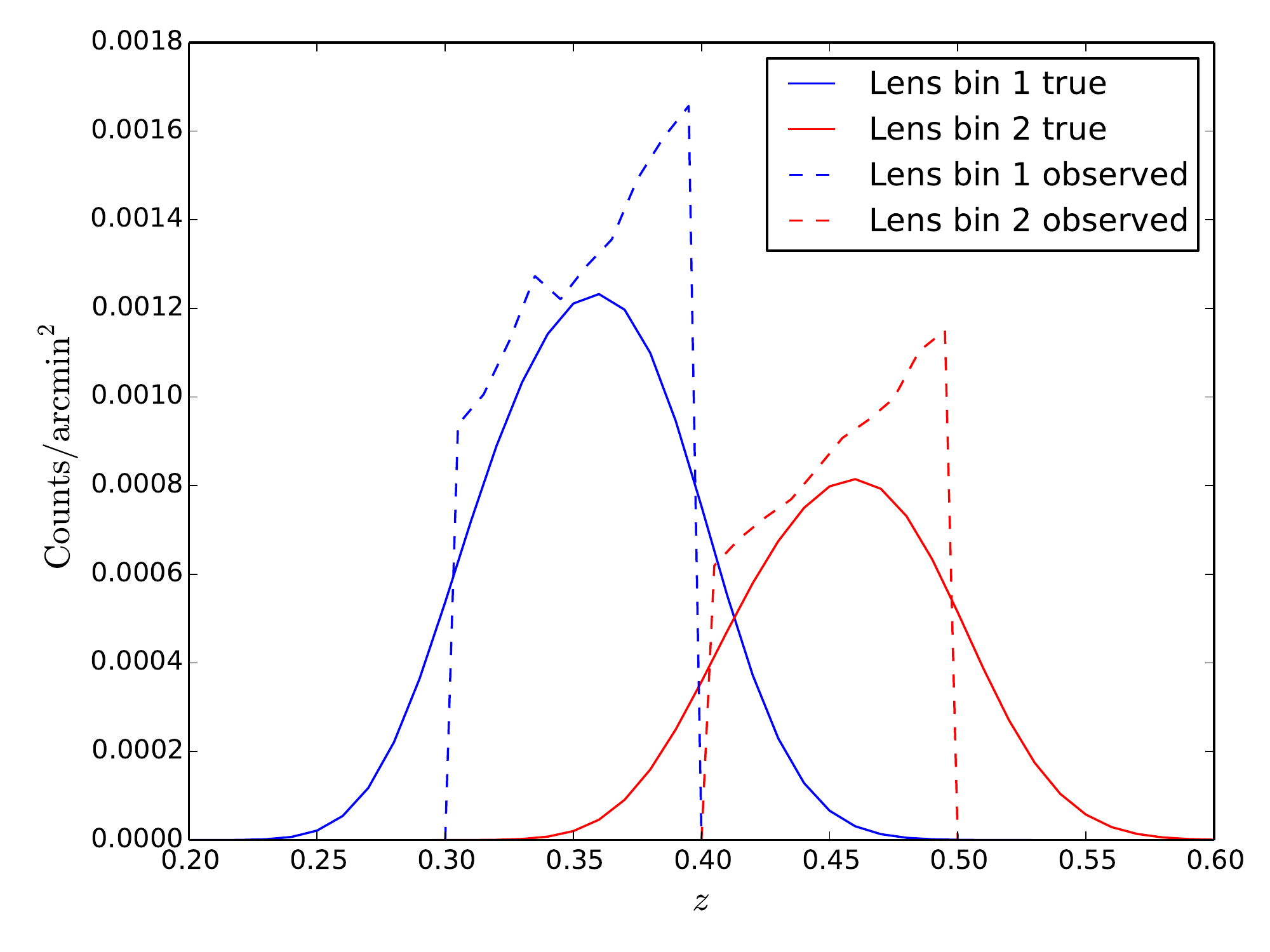}
\includegraphics[width=0.49\textwidth]{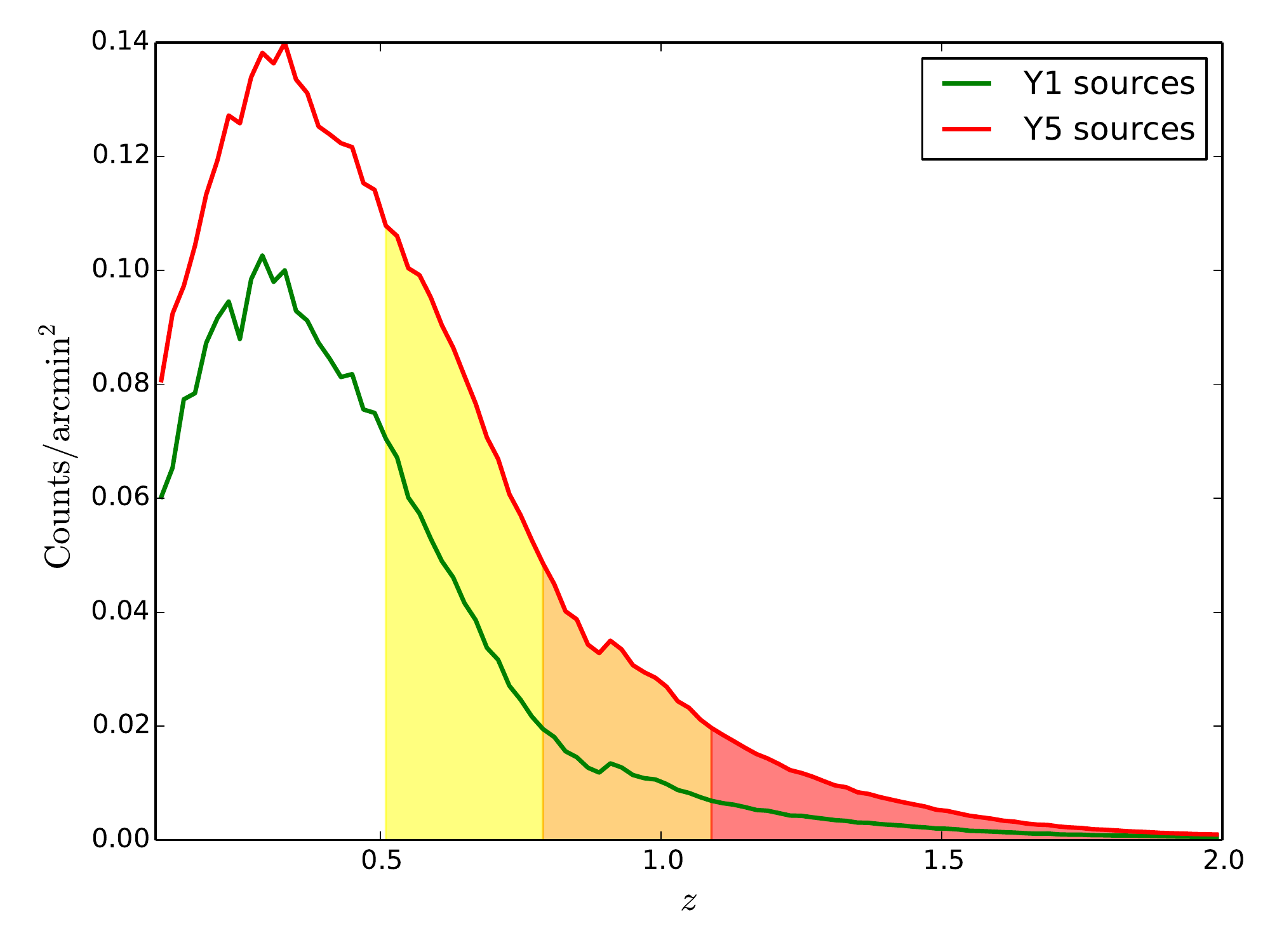}
\caption{The lens redshift distributions (left) and source redshift distributions (right) are presented. For the lenses, we show the measured redshift distribution (dashed) and the deduced true redshift distribution (solid) for the two lens bins of our mock catalog. For the sources, we show the Y1 and Y5 redshift distributions, with color-filled regions indicating the three source bins used.}
\label{fig:zdist}
\end{figure}

The source sample is selected from the DES shear mock catalog by additionally imposing $m_{\mathrm i} < 23.0$ for the Y1 source sample, and $m_{\mathrm i} < 23.5$ for the Y5 source sample to model the different depth of these two survey stages (Huan Lin, private communication). The resulting source catalog has an effective source density of $4.34$ galaxies/$\mr{arcmin}^2$ ($2.70$ galaxies/$\mr{arcmin}^2$) for Y5 (Y1). While the $m_{\mathrm i} < 23.5$ is shallower than the nominal survey depth of $m_{\mathrm i}\sim24.0$, the resulting source galaxy density for Y5 is comparable to that of current shear catalogs for the SVA data \citep{Jarvis15}.
We divide these background sources into three source redshift bins, $0.5 < z^1_s < 0.8$, $0.8< z^2_s < 1.1$, and $1.1 < z^3_s < 2.0$. Figure \ref{fig:zdist} shows the resulting redshift distributions of lens and source galaxies.
The source tomography bins contain $n^j_{\rm{gal}} =\{1.25,0.46,0.28\}$ galaxies/$\mr{arcmin}^2$ for our DES Y5 model, and $n^j_{\rm{gal}} =\{0.65,0.18,0.11\}$ galaxies/$\mr{arcmin}^2$ for our DES Y1 model.

\subsection{Measurement Vector}
\label{sec:dv}
For our simulated likelihood analysis, we generate a measurement vector from our modeling framework assuming a set of fiducial default values for parameters. That is, we use the output of our prediction codes for $\gamma_\mr{t}(\theta)$ and $w(\theta)$ under the fiducial default parameter settings as our measurement vector. We will refer to this as the simulated measurement vector. This, by construction, ensures that we can examine the information content of the proposed method, which is the goal of this paper, independent of discrepancies between simulations and theoretical models. 

We choose the small-scale lensing data vector to range from 1 to 6 arcminutes across 9 logarithmic bins, and the large-scale clustering data vector to range from 15 to 150 arcminutes across 10 logarithmic bins. This choice of scales separates the one-halo regime from the large-scale, linear clustering regime, and cuts out the transition and weakly non-linear clustering regimes, where the theoretical modeling uncertainty is the largest.

\subsection{Covariance Estimation}
We approximate the survey geometry of the Y1 and Y5 DES footprint as rectangles of $1000$ and $5000$ square degrees, respectively. We use the tree code \texttt{treecor} \citep{Jarvis04} to calculate $\gamma_{\mathrm t}(\theta)$ and $w(\theta)$ (using the Landy-Szalay estimator \cite{LS} with uniform random mocks for the latter), and measure the joint covariances by the bootstrap-with-oversampling method of \citet{Norberg09}, using $20$ square degree patches and an oversampling factor of 3, yielding
\begin{equation}
\mathrm{Cov}(d_i,d_j) = \frac{1}{N-1}\sum_{k=1}^{N}\left(d_i^k - \bar{d}_i\right)\left(d_j^k - \bar{d}_i\right)\,
\end{equation}
with the joint data vector $\mathbf d = \left(w(\theta_{1,...,N_{\mathrm w}}), \gamma_{\mathrm t}(\theta_{1,...,N_\gamma},z_{\mathrm s}^1), \gamma_{\mathrm t}(\theta_{1,...,N_\gamma},z_{\mathrm s}^2), \gamma_{\mathrm t}(\theta_{1,...,N_\gamma},z_{\mathrm s}^3)\right)$, $\mathbf{d}^k$ the $k$-th bootstrap realization, $N = 3 N_{\mathrm{patch}}$ the number of bootstrap samples, and $ \bar{\mathbf d}$ the mean data vector calculated as
\begin{equation}
 \bar{\mathbf d} = \frac{1}{N}\sum_{k=1}^{N} \mathbf d^k\,.
\end{equation}
We estimate the joint clustering and galaxy-galaxy lensing covariance for the two lens bins separately, and assume a block-diagonal total covariance matrix for the combination of multiple lens bins.

\begin{figure}[htbp]
\includegraphics[width=0.9\textwidth]{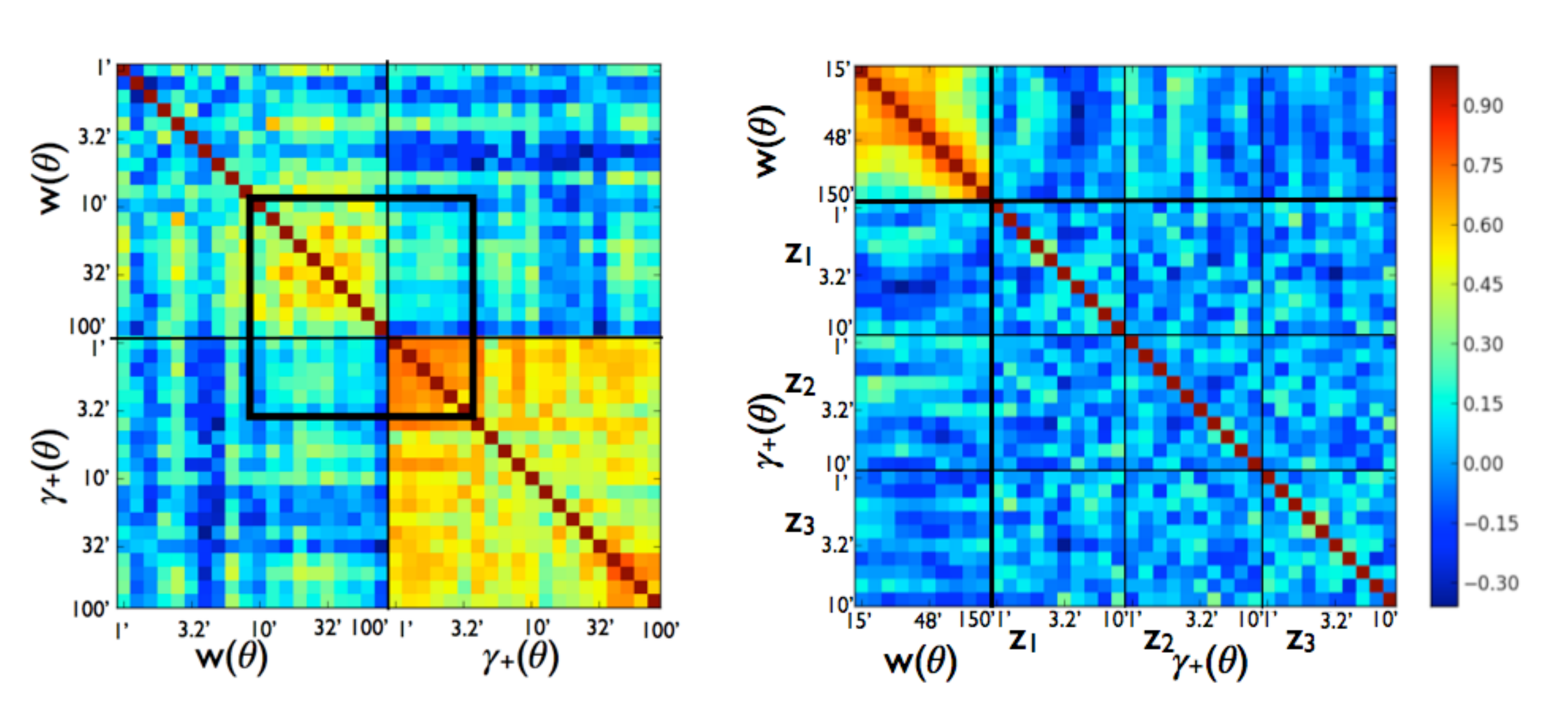}
\caption{\textit{Left}: Correlation matrix of galaxy clustering and galaxy-galaxy lensing with a single source bin for illustrative purposes. A large range of scales ($1'-100'$) is shown with a larger lens sample to reduce shot/shape noise and highlight the correlations of density modes. The black box indicates the range of scales considered in this analysis. \textit{Right}: Correlation matrix of galaxy clustering and tomographic galaxy-galaxy lensing for the DES Y5 $0.3<z<0.4$ lens sample and range of scales considered in this analysis (cf.~\ref{sec:dv}). The panels marked as $z_j$ correspond to the tangential shear measurement vector at the jth source redshift bin.}
\label{fig:Cor}
\end{figure}

The left panel in Figure~\ref{fig:Cor} shows the correlation matrix of clustering and galaxy-galaxy lensing over a larger range of scales than those considered in this analysis to illustrate the correlation between scales and probes. The black box indicates the range of scales considered in the analysis. Note that this covariance matrix is based on a larger lens sample in order to reduce statistical noise and highlight the underlying correlations due the correlation of density modes and the fact that galaxy-galaxy lensing and clustering both probe the underlying matter density field. In the right panel, we show the actual correlation matrix for our Y5 data vector in a single lens bin, with the tangential shear measurements from the three source bins marked as $z_i$. We observe reduced off-diagonal covariances, as shape noise and shot noise (respectively) are higher for the tomographic galaxy-galaxy lensing and the clustering of the lens galaxy sample used in our analysis.

Note that while we choose the mock lens galaxy samples used in the covariance estimation to be similar in mass range and number density to the fiducial lens galaxy sample used for generating the measurement vector, the match is not exact. In order to adjust for the difference in signal strength to leading order, we rescale the clustering auto-covariance, clustering -- galaxy-galaxy lensing cross-covariance, and galaxy-galaxy lensing auto-covariance by their respective scaling with galaxy bias, i.e. by $(b_{\rm{fid}}/b_{\rm{mock}})^4$, $(b_{\rm{fid}}/b_{\rm{mock}})^3$, and $(b_{\rm{fid}}/b_{\rm{mock}})^2$ respectively. Here, $b_\mr{fid}$ is the galaxy bias calculated for the synthetic measurement vector, and $b_\mr{mock}$ is the galaxy bias measured from the mock data. This covariance rescaling is equivalent to performing an analysis using the original covariance with rescaled HOD-derived data vector $\left(w,\gamma_{\rm t}\right)\rightarrow \left((b_{\rm{mock}}/b_{\rm{fid}})^2 w, (b_{\rm{mock}}/b_{\rm{fid}})\gamma_{\rm t}\right)$, and does not change the shot noise level. The latter is difficult to adjust in real-space covariances as (due to a mixed cosmic variance and shot noise term) it affects all covariance elements differently \cite{Eifler09}. Since the number densities corresponding to our fiducial HOD parameters for the lens sample are higher than the number densities of the mock samples, this is a conservative rescaling and may overestimate statistical errors.

\section{Likelihood Analysis}
\label{sec:analysis}
\subsection{Overview}
With our prediction from Section \ref{sec:impl} and mock data from Section \ref{sec:data}, we perform a Markov Chain Monte Carlo (MCMC) likelihood analysis to forecast how well this analysis can constrain model parameters under various data stages of DES. 
To generate the simulated measurement vector used in these analyses, we must define a set of fiducial default values for model parameters. These fiducial defaults represent our best-guess estimates for the model parameters characterizing actual DES data. In addition, for the likelihood analysis, a set of priors for these parameters must be assumed. Priors allow us to include information outside of our pipeline, either from DES or from external results, that strengthens our constraining power. It is important to note that our priors on the HOD and systematic effects parameters represent the constraining power we expect to obtain with DES data outside of our pipeline, even when we benchmark our estimates from external results where we lack existing analyses of DES data. Also, since we use a simulated measurement vector without random errors here, i.e. without introducing 
further random fluctuations to the output from the prediction code, we focus on investigating the constraining power and degeneracies implied by the obtained constraints when we look at the final results, using the central values only as reference points. 

Below, we first detail the full parametrization of our likelihood analysis employed for the simulated Y1 and Y5 analyses.

\subsection{Parameter Space}
\label{subsec:priors}

\begin{table}[htbp]
\begin{tabular*}{\columnwidth}{@{\extracolsep{\stretch{1}}}*{5}{c}@{}}
\hline
\hline
\noalign{\vskip 1mm}
Sector                          & Parameter         & Fiducial Default     & Prior width (1-$\sigma$) & Source \\
\hline
\noalign{\vskip 1mm}
\multirow{3}{*}{Cosmology}      & $\Omega_M$        & 0.314                & \multirow{3}{*}{Planck likelihoods} & \multirow{3}{*}{\citet{planckcmb}} \\
                                & $h$               & 0.673                &  &  \\
                                & $A_s$             & 2.15$\times 10^{-9}$ &  &  \\
\hline
\noalign{\vskip 1mm}                              
\multirow{4}{*}{HOD}            & $\log M_\mr{min}$ & 12.36, 12.33         & 0.09               & \multirow{4}{*}{\citet{Coupon:2011mz}}       \\
                                & $\log M_1$        & 13.69, 13.58         & 0.05               &        \\
                                & $\sigma_{\log M}$ & 0.32, 0.30           & 0.15               &        \\
                                & $\alpha$          & 1.28, 1.37           & 0.05               &        \\
\hline
\noalign{\vskip 1mm}
\multirow{2}{*}{Lens Photo-z}   & $\sigma_{zL}$     & 0.02, 0.02           & 0.01               & \multirow{2}{*}{\citet{redmagic}}      \\
                                & $b_{zL}$          & 0, 0                 & 0.01               &        \\
\hline
\noalign{\vskip 1mm}
\multirow{2}{*}{Source Photo-z} & $\sigma_{zS}$     & 0.08                 & 0.01               & \multirow{2}{*}{\citet{carlesphotoz,Bonnett15}}       \\
                                & $b_{zS}$          & 0                    & 0.01               &        \\
\hline
\noalign{\vskip 1mm}
Shear Calibration               & $m_\gamma$        & 0                    &  0.02              & \citet{Jarvis15,Clampitt15}       \\
\hline
\noalign{\vskip 1mm}
Growth Scaling                  & $A_i$             & 1, 1                 &  Flat [0.5, 2.0]   & N/A       \\
\hline
\hline
\end{tabular*}
\caption{List of parameters with their fiducial defaults and 1-$\sigma$ prior widths presented with the respective sources from which we draw these values. Entries with a pair of values represent parameters that vary between the two lens bins, while entries with a single value represent parameters that are global for both bins. Note that the prior widths are default settings for the conservative Y5 analysis; for analyses with different assumptions, subsets of parameter widths are varied as stated below. All mass values are units of $M_\odot / h$.}
\label{tab:params}
\end{table}

The mock Y1/Y5 survey setup described in \ref{subsec:y1y5data} yields a 20-dimensional parameter space. These parameters can largely be classified into cosmological, HOD, systematic effects, and growth scaling parameters. Here, we discuss how we set parameter defaults and priors for each parameter category, with references to relevant DES analyses on the SVA1 data as well as external results serving as benchmarks. Table \ref{tab:params} lists the numerical values for parameter defaults and priors in detail.

\textbf{Cosmology} -- For cosmological parameters, we combine the Planck likelihood \citep{planckcmb} with the likelihood that emerges from our pipeline, thereby enforcing Planck priors. Accordingly, our fiducial model takes Planck best fit parameters as defaults. With the addition of growth scaling parameters, our model apparently has three parameters ($A_s$, $b_g$, and $A_i$) that shift the overall clustering strength for each lens bin. However, galaxy bias is not treated as a free parameter, but rather as a function of the halo and galaxy mass distribution (Equation ~\ref{eq:bg}), and with the constraints on $A_s$ from Planck, we are able to constrain $A_i$ independently as initially suggested by \citet{Yoo:2012}. 

\textbf{HOD} -- HOD priors represent the additional constraining power on HOD parameters that we expect to obtain from information not used by our current setup. For example, since our analysis does not use small-scale galaxy clustering, we can imagine including HOD constraints from an independent small-scale galaxy clustering analysis as HOD priors. Or, if we later include small-scale galaxy clustering in our analysis, the expected strengthening of HOD constraints can be emulated by HOD priors in our current setup. Since an independent, HOD-focused analysis is yet to be performed on DES data, we use the results of the CFHTLS-Wide survey~\citep{Coupon:2011mz} as a benchmark for the eventual DES HOD constraints. For fiducial defaults, we adopt the CFHTLS best-fit HOD parameters with a comparable luminosity and redshift selection. For priors, we consider two primary sets of assumptions. The first set, which we refer to as conservative, assumes that DES Y5 data will yield HOD constraints equivalent to the CFHTLS results, and use the CFHTLS 1-$\sigma$ uncertainties as default widths of Gaussian priors on the HOD parameters in the simulated Y5 analysis, as detailed in Table \ref{tab:params}. In the simulated Y1 analysis, we double the default prior widths for $\log M_\mr{min}$, $\log M_1$, and $\alpha$ to reflect the relatively smaller sky coverage and shallower depth of the Y1 data stage. The second set, which we refer to as optimistic, assumes that DES Y1 and Y5 HOD constraints will scale with their increased sky coverages compared to CFHTLS, and uses HOD prior widths decreased by factors of $\left(f_\mr{sky,Y1}/f_\mr{sky,CFHTLS}\right)^{1/2}$ and $\left(f_\mr{sky,Y5}/f_\mr{sky,CFHTLS}\right)^{1/2}$ for Y1 and Y5, respectively. These factors are roughly 2.7 and 6.1 for Y1 and Y5. One of the key issues in our analysis is how much information is needed about the HOD parameters in the quest to constrain the cosmological parameters $A_i$, and to study the effect of these priors on our eventual constraining power, we also carry out several conservative Y1 analyses where the the widths for $M_\mr{min}$, $M_1$ and $\alpha$ are loosened by factors of 1.5, 2.5, 3.5, and 5. 

\textbf{Systematic Effects} -- By systematic effects parameters, we refer to the lens photo-z, source photo-z, and the multiplicative shear calibration parameters. We expect to understand the extent of photo-z errors present in DES catalogs from studies of spectroscopic subsamples and simulations, and this information can be incorporated into this analysis through photo-z priors. The lens photo-z modeling and priors adopted above are realistic for an LRG galaxy sample, and we anticipate the first application of our data to use DES redMaGiC~\citep{redmagic} galaxies as the lens sample. The redMaGic galaxy sample is selected by fitting every galaxy to a red sequence template and establishing chi-squared cuts to enforce a constant comoving spatial density of galaxies over redshift, which by design allows for the selected galaxies to have tight and well-behaved (Gaussian) photo-z constraints. The ``pessimistic'' redMaGiC photo-z estimates are reported as $\sigma_{zL}=0.015$ and $b_{zL}=0$ with $\sim 1\%$ catastrophic redshift failure rate, and we use conservative values of $\sigma_{zL}=0.02$ and $b_{zL}=0$ as our defaults. For the photo-z precision of source galaxies, early photo-z results in DES data~\citep{carlesphotoz} suggests $\sigma_{zS} = 0.08$ and $b_{zS}=0$, which we use as defaults. We adopt Gaussian priors of width $0.01$ for these four parameters, allowing both the bias and the variance of the photometric redshift estimates to be determined by the data, subject to modest priors on their ultimate values. In addition, we note that while tests of consistency between the lensing from different source redshift bins shows no discrepancies in the SVA1 data (\citet{Clampitt15}), such tests do not account for an overall multiplicative bias that would affect all source bins equally. This multiplicative shear calibration parameter, $m_\gamma$, is measured in \citet{Jarvis15} and found to be less than 2\%. Thus, we assume $m_\gamma=0$ as our fiducial default, and introduce a 0.02 (2\%) Gaussian prior on this parameter. Finally, similar to the HOD priors, we carry out a mock Y1 analysis where the prior widths for systematic effects parameters are widened by a factor of 2.5 to gauge how they affect our final constraining power on $A_i$. This exercise is extended to a number of optimistic Y5 analyses with systematic effects prior widths of 0.5, 2, and 4 times the default width, as the optimistic Y5 scenario is expected to exhibit the strongest impact from systematic effects parameters.

\textbf{Galaxy abundance priors} -- Galaxy abundance, or the total number of galaxies in the survey, is calculated as
\be
N_g = \Omega_s \int dz \frac{\chi^2}{H(z)}\bar{n}_g(z)
\ee
where $\Omega_s$ is the solid angle subtended by the survey.
The calculation of $\bar{n}_g$, from Equation \ref{eq:ngal}, has a different dependence on the mass function and the HOD than does the galaxy bias $b_g$ from Eq. \ref{eq:bg}, so simply counting the number of galaxies in the survey will provide an additional constraint on HOD parameters. In particular, the number of galaxies in the survey 
breaks a problematic degeneracy between $\alpha$ and $A_i$. To implement galaxy abundance priors in our simulated examples, we will assume a generic 10\% scatter in $N_g$ for Y1 and 5\% for Y5, and adopt corresponding Gaussian likelihoods into the analysis. Note that under the most ideal circumstances, there is only the Poisson and sample variance uncertainties on $N_g$. However, since galaxy selection is diluted by uncertainties in the photo-z and the mass-luminosity relation, we choose to adopt these conservative prior widths.

\subsection{Running and Verifying Chains}

To implement the MCMC, we use \texttt{CosmoSIS}~\citep{Zuntz:2014csq}, a modular parameter estimation framework. 
For MCMC sampling, we make use of the \texttt{emcee} sampler~\citep{ForemanMackey:2012ig}, an implementation of the affine-invariant MCMC ensemble sampler discussed in \citet{Goodman2010}, using 190 walkers. Each ensemble iteration in our MCMC thus consists of 190 samples, one from each walker. In order to ensure that our chains are properly drawing independent samples from the likelihood space, we utilize measurements of the integrated autocorrelation time $\tau$ as a criterion for testing convergence. Denoting the mean value of parameter $p_i$ in the t-th ensemble iteration as $\hat p_i(t)$, the autocorrelation function $C_i(T)$ for that parameter with ensemble iteration lag $T$ is given by
\be
C_i(T) = \ensav{(\hat p_i(t+T)-\ensav{\hat p_i})(\hat p_i(t)-\ensav{p_i})}.
\ee
The autocorrelation function is commonly normalized as
\be
\rho_i(T) = C_i(T)/C_i(0),
\ee
which then yields the integrated autocorrelation time $\tau_i$ as
\be
\tau_i = \frac{1}{2}+\sum_{T=1}^{T_\mr{max}}\rho_i(T).
\ee
For a properly converged chain, $\tau_i$ reaches an asymptotic value and is stable with respect to $T_\mr{max}$, and this behavior then can be used as a heuristic signal for convergence. For our chains, stabilized $\tau_i$ values for different parameters range from 40 to 130 ensemble iterations. As suggested in \citep{ForemanMackey:2012ig}, we then consider the first few (around 10) $\tau_i$ ensemble iterations as burn-in, and choose to discard the first 1000 ensemble iterations. Parameter estimation is then performed on the following 900 ensemble iterations, consisting of 171,000 samples. In addition, we check that the acceptance fraction observed in our chain is stabilized to a reasonable value for the chosen region.

\section{Results}
\label{sec:results}

This section presents the results from the likelihood analyses described in Section \ref{sec:analysis}, revolving around ``triangle'' plots of 1-D and 2-D constraints of model parameters. In visualizing our results, we use a modified version of the \texttt{triangle}~\citep{triangle} Python package. In each subsection below, we will focus mostly on particular subsets of parameters. 

Large plots encompassing full parameter sets are pushed off to Appendix \ref{sec:figs}. In particular, Figs. \ref{fig:contours} and \ref{fig:contours2} are our forecast parameter constraints from the default simulated Y1 and Y5 likelihood analyses. As discussed in Section \ref{sec:analysis}, these results represent our conservative and optimistic estimates at the eventual DES constraining power on the growth function for the respective data stages. The takeaway is that parameter constraints are very well centered with respect to their true values, an indication that the 20-parameter MCMC is working well. Note that we are not showing constraints on the cosmological parameters, as these are largely dominated by Planck priors. Therefore, we suppress these columns but come away with the knowledge that for the cosmological parameters that we consider -- $\Omega_M$, $h$, and $A_s$ -- we expect CMB constraints to be dominant over constraints from combining small scale lensing and large scale clustering. Also, as the two lens bins show very similar parameter behaviors, we only show contours for the first lens bin and simply tabulate results for the second lens bin.

In all triangle plots shown below, the panels along the diagonal correspond to the 1-D probability distributions for each parameter with dotted vertical lines at the 16th and 84th percentiles, while the off-diagonal panels show the 2-D Gaussian 1-$\sigma$ confidence contours for the corresponding pair of parameters. The light blue lines and squares represent the fiducial default parameter values used to generate the simulated measurement vector, i.e. the ``true'' parameter values. Numerical values for marginalized 1-$\sigma$ bounds are listed in Table \ref{tab:bounds}.

\subsection{Default Conservative and Optimistic Results}

\begin{figure}[h]
\includegraphics[width=0.5\textwidth]{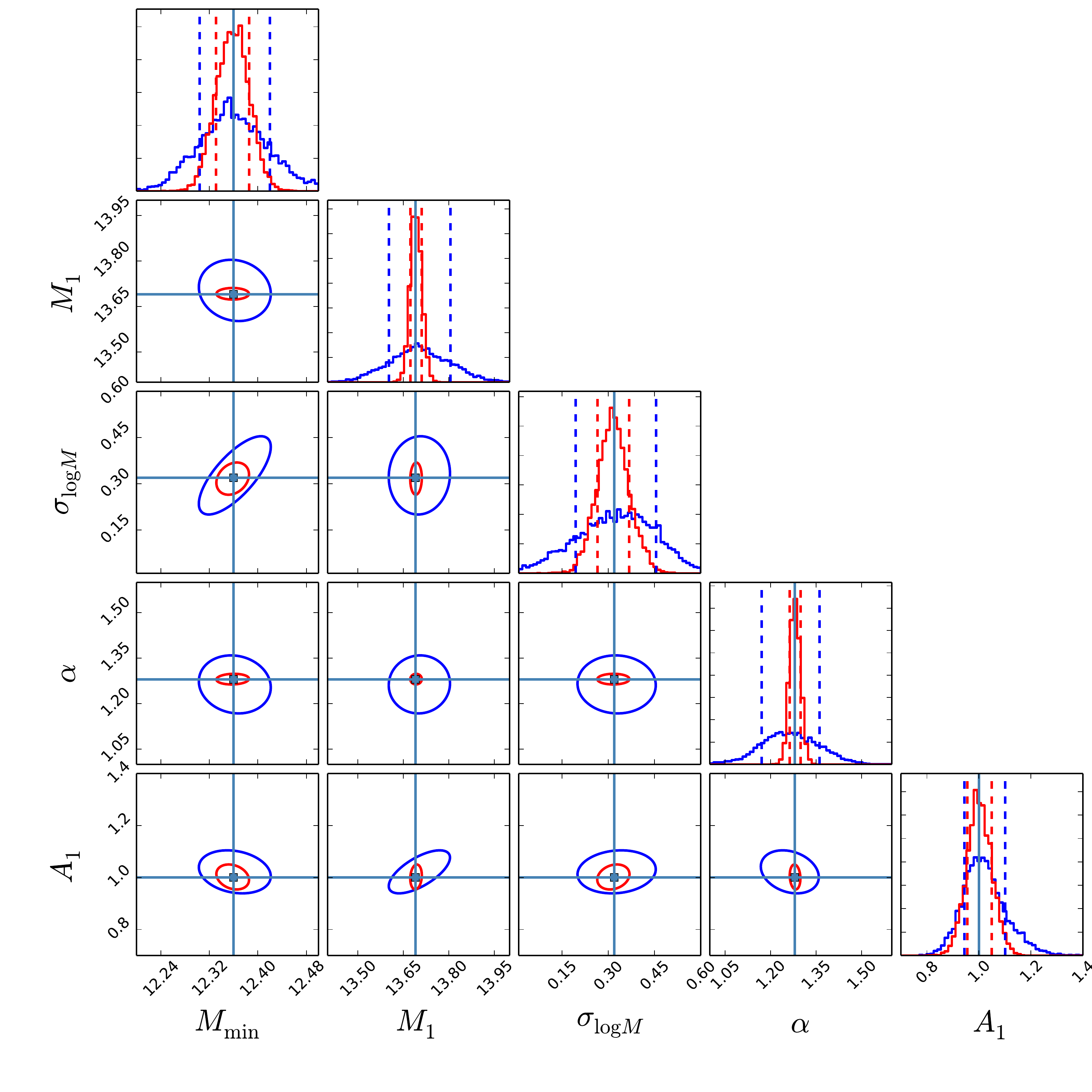}
\caption{Forecast constraints on the HOD and growth scaling parameters from the conservative (blue) and optimistic (red) Y1 analyses. Diagonal blocks represent marginalized 1-D parameter constraints, and off-diagonal blocks represent 2-D 1-$\sigma$ confidence ellipses for corresponding parameters.}
\label{fig:y1_consopt_hod}
\end{figure}

Let us begin with results under the two default -- conservative and optimistic -- assumptions for the simulated Y1 analysis. In Fig. \ref{fig:y1_consopt_hod}, we present our forecast parameter constraints on the HOD and the growth scaling parameters from those two analyses. A key issue for our study is the extent of the correlation between $A_i$, representing the amplitude of matter fluctuations in the lens bins, and the HOD parameters. If there were no degeneracy, then the analysis could be carried out without any dependence on HOD modeling. Fig. \ref{fig:y1_consopt_hod} shows that this is not the case, i.e. that the HOD parameters are correlated with $A_i$. In particular, $A_i$ are quite degenerate with the two parameters that quantify the satellite galaxy abundance, $M_\mr{1}$ and $\alpha$. 

The difference between the two analyses is in widths of HOD priors, where the conservative analysis assumes widths twice as large as CFHTLS constraints, while the optimistic analysis assumes widths smaller by a factor of $\left(f_\mr{sky,Y1}/f_\mr{sky,CFHTLS}\right)^{1/2}$ (roughly 2.7), resulting in a net difference in HOD widths by a factor of 5.4. For central HOD parameters $M_\mr{min}$ and $\sigma_\mr{\log M}$, we observe that the difference between the conservative and optimistic constraints is smaller than the difference in the prior widths. For satellite HOD parameters $M_\mr{1}$ and $\alpha$, we observe the difference in constraints closely following the difference in the prior widths, indicating that these constraints are largely prior-driven. From the two analyses, we project 7.9\% (conservative) and 4.7\% (optimistic) 1-$\sigma$ error bars on $A_1$.

\begin{figure}[h]
\includegraphics[width=0.5\textwidth]{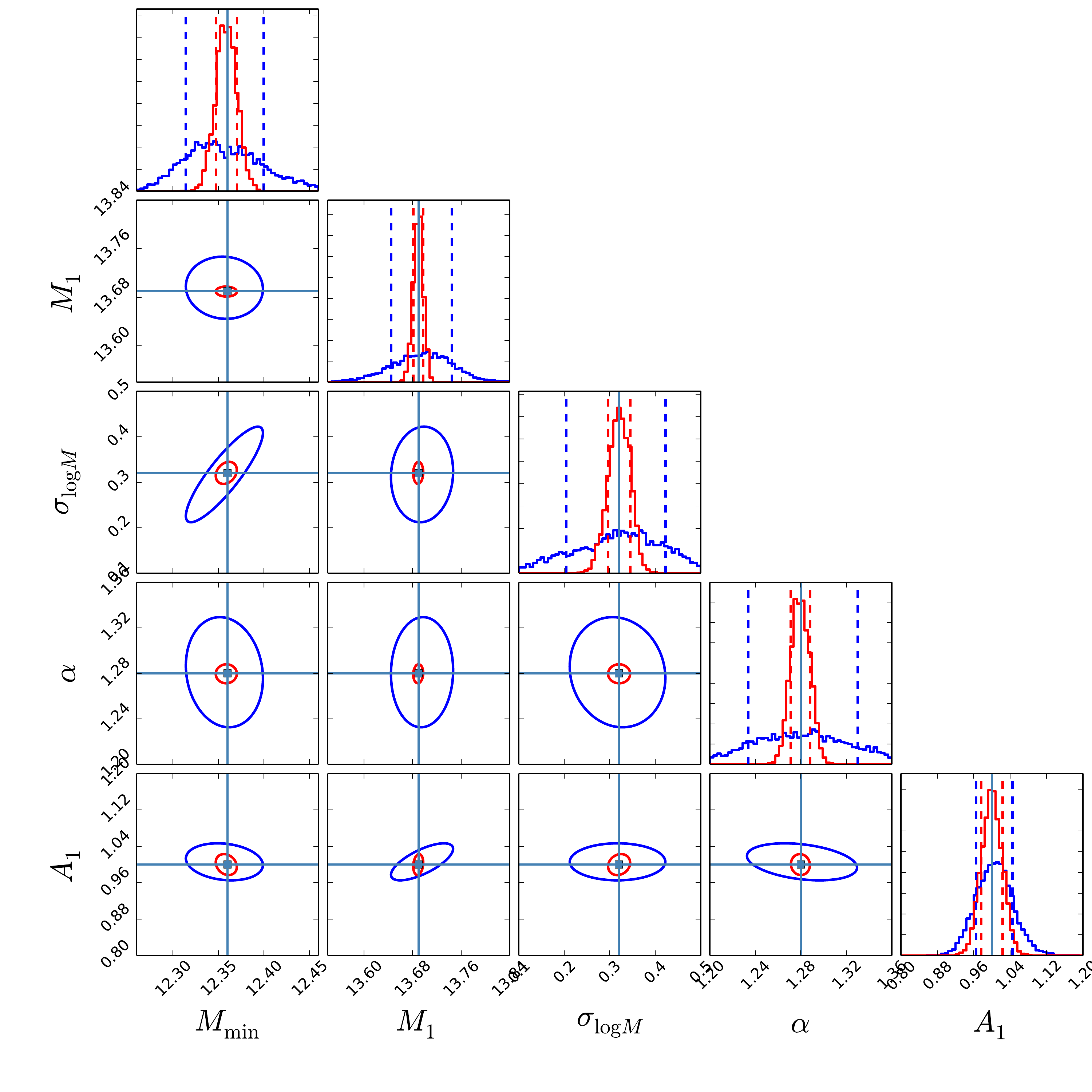}
\caption{Forecast constraints on the HOD and growth scaling parameters from the conservative (blue) and optimistic (red) Y5 analyses. Diagonal blocks represent marginalized 1-D parameter constraints, and off-diagonal blocks represent 2-D 1-$\sigma$ confidence ellipses for corresponding parameters.}
\label{fig:y5_consopt_hod}
\end{figure}

In Fig. \ref{fig:y5_consopt_hod}, we compare the results from the conservative and optimistic Y5 analyses. The conservative analysis assumes widths equivalent to CFHTLS constraints, while the optimistic analysis assumes widths smaller by a factor of $\left(f_\mr{sky,Y5}/f_\mr{sky,CFHTLS}\right)^{1/2}$ or roughly 6.1. We observe similar parameter behaviors as in the Y1 counterparts, and project 3.9\% (conservative) and 2.3\% (optimistic) 1-$\sigma$ error bars on $A_1$.

\begin{figure}[h]
\includegraphics[width=0.5\textwidth]{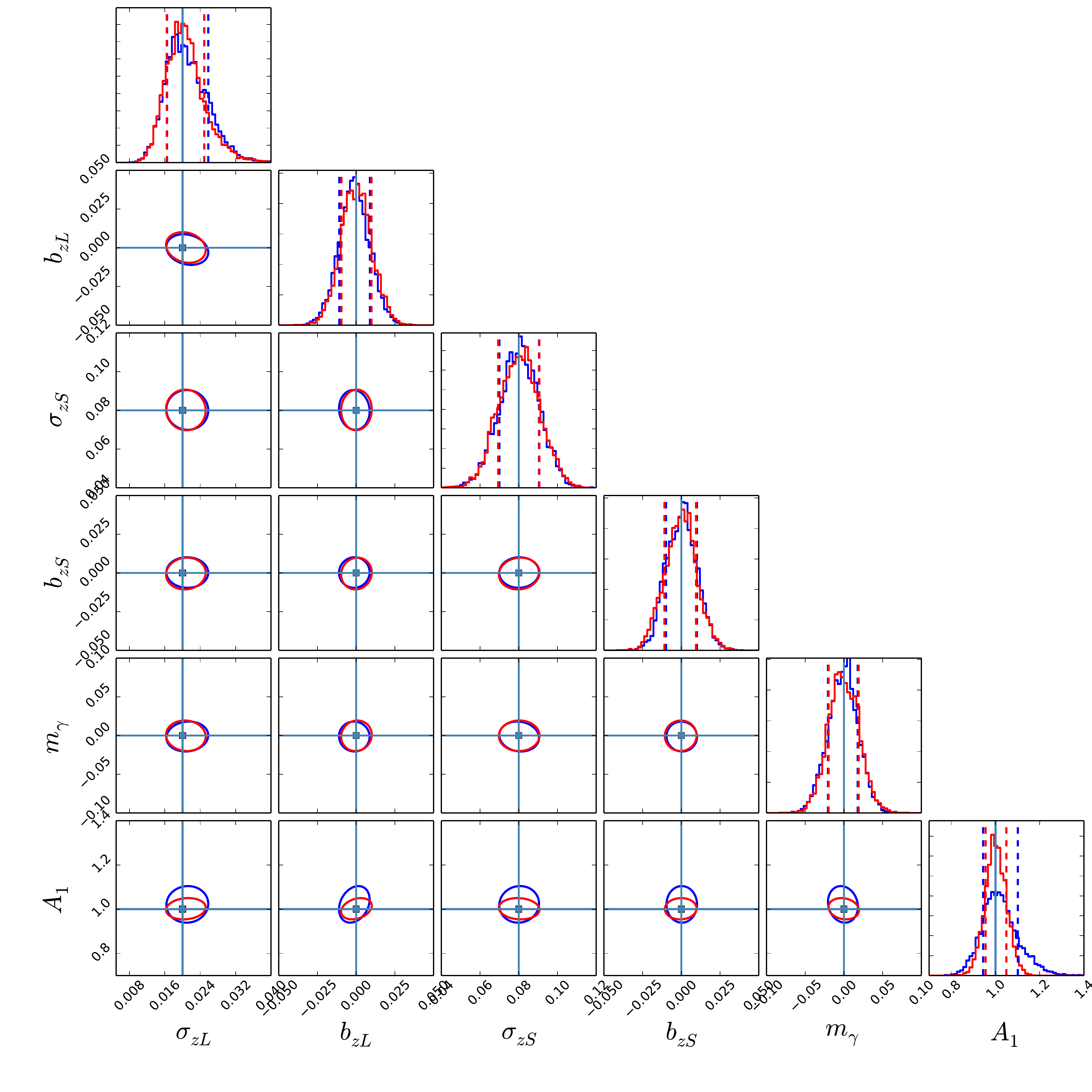}
\caption{Forecast constraints on the systematic effects and growth scaling parameters from the conservative (blue) and optimistic (red) Y1 analyses. Diagonal blocks represent marginalized 1-D parameter constraints, and off-diagonal blocks represent 2-D 1-$\sigma$ confidence ellipses for corresponding parameters.}
\label{fig:y1_consopt_sys}
\end{figure}

\begin{figure}[h]
\includegraphics[width=0.5\textwidth]{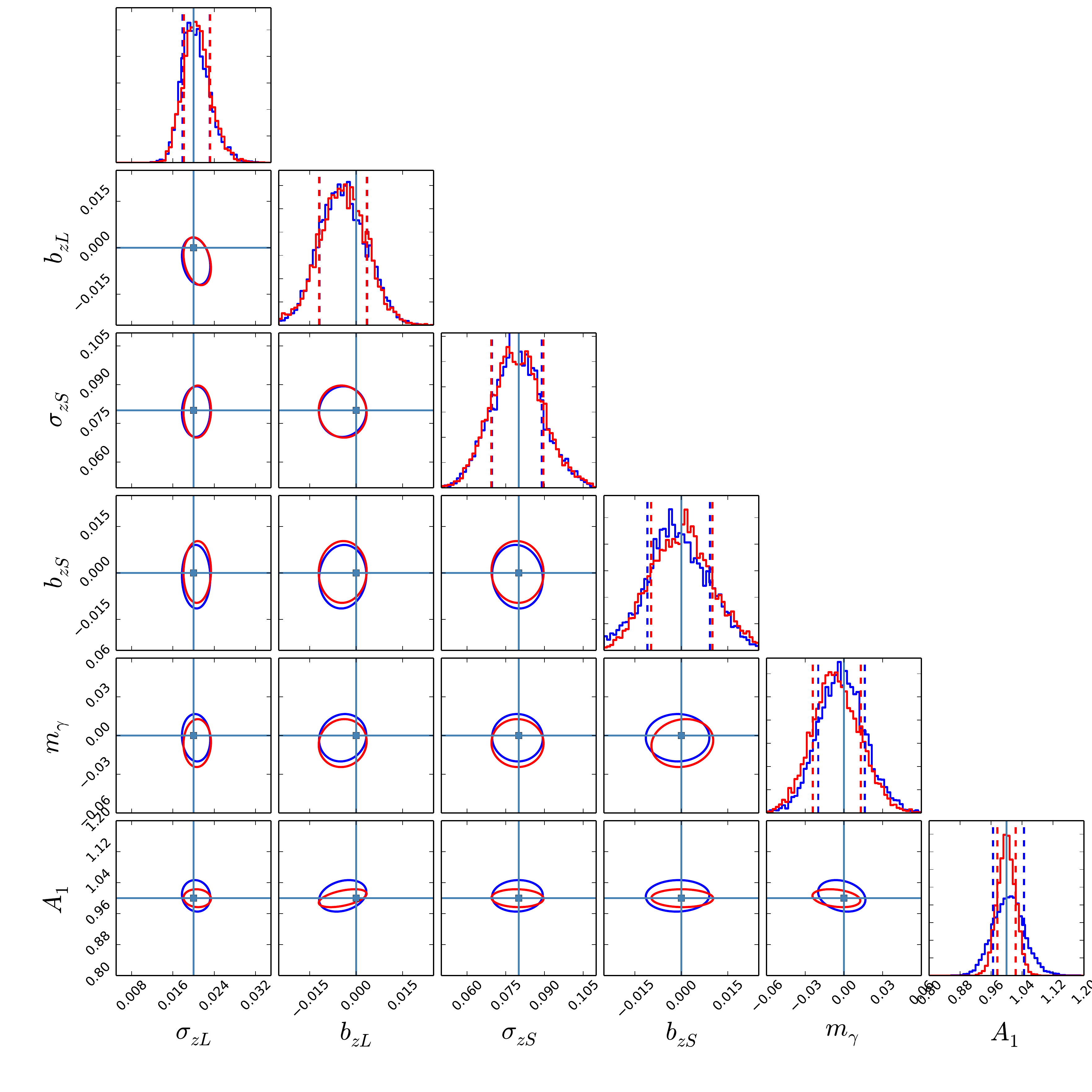}
\caption{Forecast constraints on the systematic effects and growth scaling parameters from the conservative (blue) and optimistic (red) Y5 analyses. Diagonal blocks represent marginalized 1-D parameter constraints, and off-diagonal blocks represent 2-D 1-$\sigma$ confidence ellipses for corresponding parameters.}
\label{fig:y5_consopt_sys}
\end{figure}

Let us now turn our attention to the systematic effects parameters. In Fig. \ref{fig:y1_consopt_sys}, we present our forecast parameter constraints on the systematic effects and the growth scaling parameters from the conservative and optimistic Y1 analyses. As opposed to the HOD parameters, Fig. \ref{fig:y1_consopt_sys} shows that there are no notable degeneracies between systematic effects parameters and our parameter of interest $A_i$. In Fig. \ref{fig:y5_consopt_sys}, we show the conservative and optimistic Y5 constraints for systematic effects and growth scaling parameters, and observe identical parameter degeneracies. Also, note that for both results how small the difference in constraints is between conservative and optimistic results. This implies that changing HOD priors has only a marginal effect on systematic effects constraints, i.e. that HOD and systematic effects parameters show little degeneracy between them. This is also observable in Figs. \ref{fig:contours} and \ref{fig:contours2}.

To further explore the effect of degeneracies between different parameter subsets and $A_i$, as well as to gauge how much constraining power is coming from the analysis itself as opposed to assumed priors, we perform a series of exercises in relaxing the priors for different parameter subsets.

\subsection{Relaxing HOD Priors}

\begin{figure}[htpb]
\includegraphics[width=0.6\textwidth]{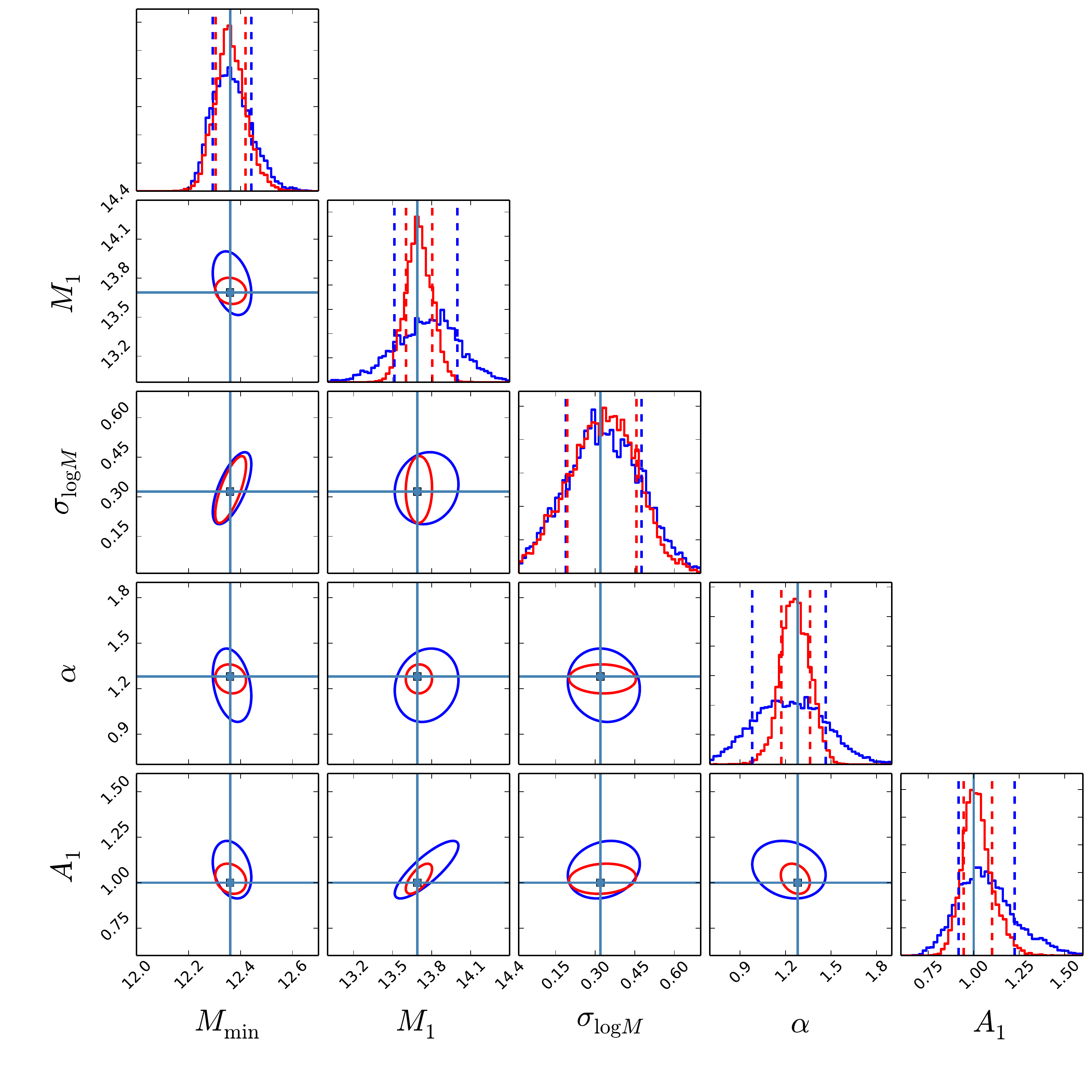}
\caption{Forecast constraints on the HOD and growth scaling parameters from the default conservative (red) and relaxed HOD priors (blue) Y1 analyses. The relaxed HOD priors are a factor of 2.5 weaker than the default priors.}
\label{fig:y1hod}
\end{figure}

To get a sense of how important the priors on the HOD parameters are, we perform a ``relaxed HOD priors'' simulated Y1 analysis, where the width of HOD priors are widened by a factor of 2.5 compared to the conservative default widths.
In Figure \ref{fig:y1hod}, we compare the conservative Y1 constraints for HOD and growth scaling parameters, as presented in Figure \ref{fig:y1_consopt_hod}, against Y1 constraints with relaxed HOD priors. Our constraining power on $M_\mr{min}$ is only mildly degraded despite the relaxed priors, implying that our analysis constrains $M_\mr{min}$ largely by itself without relying on external priors. However, $M_1$ and $\alpha$ are not as well constrained by the new, wider priors, implying that the priors are driving our constraining power on these parameters. Since these are the parameters that are most degenerate with $A_1$, it is not surprising that the constraint on $A_1$ loosens by a factor of 2, with the 7.9\% error bar on $A_1$ using nominal priors degraded to 16.1\% with the looser priors.

\begin{figure}[htpb]
\includegraphics[width=0.6\textwidth]{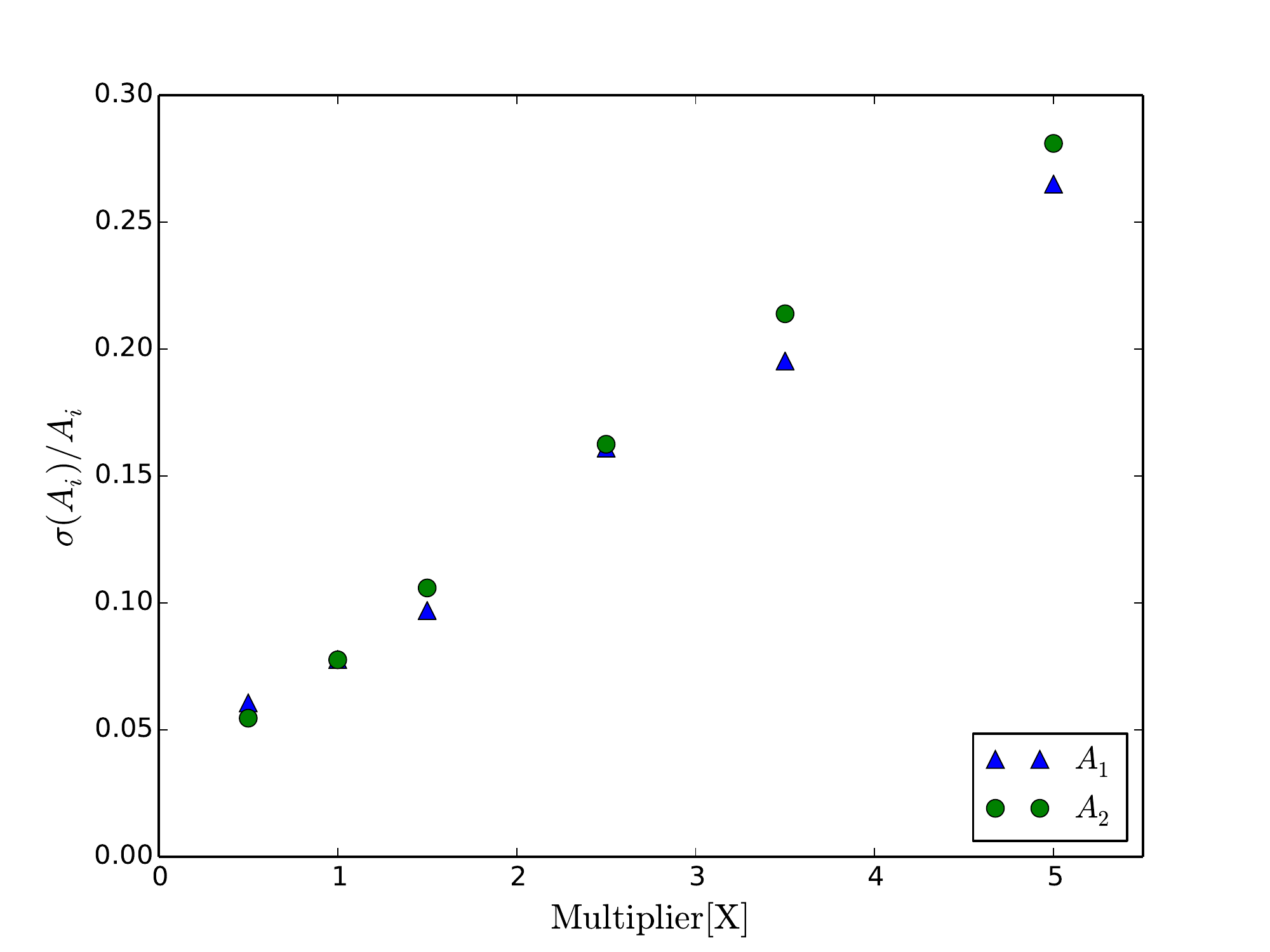}
\caption{Fractional 1-$\sigma$ uncertainty on $A_i$ for different Y1 analyses, plotted against HOD prior widths as a factor of the default conservative widths.}
\label{fig:hodeffect}
\end{figure}

In Fig.~\ref{fig:hodeffect}, we go further to quantify the extent to which $M_\mr{1}$ and $\alpha$ impact $A_i$: we compare our final constraints on the growth scaling parameters $A_{1,2}$ from a number of different Y1 analyses as a function of the prior width on the HOD parameters. In addition to the default conservative and 2.5$\times$ widths, we consider prior widths equal to 0.5, 1.5, 3.5, and 5 times the default conservative width. The minimum mass $M_\mr{min}$ is relatively well constrained by the data regardless of priors, so the main effect of varying HOD priors is on the satellite HOD parameters. The result we observe, as presented in Fig. \ref{fig:hodeffect}, is a linear relationship between the prior widths and the constraints on $A_{1,2}$, which confirms the strong degeneracy between satellite HOD parameters and $A_{1,2}$. The lessons from Fig. \ref{fig:hodeffect} are straightforward: the constraints on the cosmological parameters of interest will be limited by our ability to constrain the satellite HOD parameters using other measurements.

\subsection{Relaxing Systematic Effects Priors}

\begin{figure}[htbp]
\centering
\includegraphics[width=0.6\textwidth]{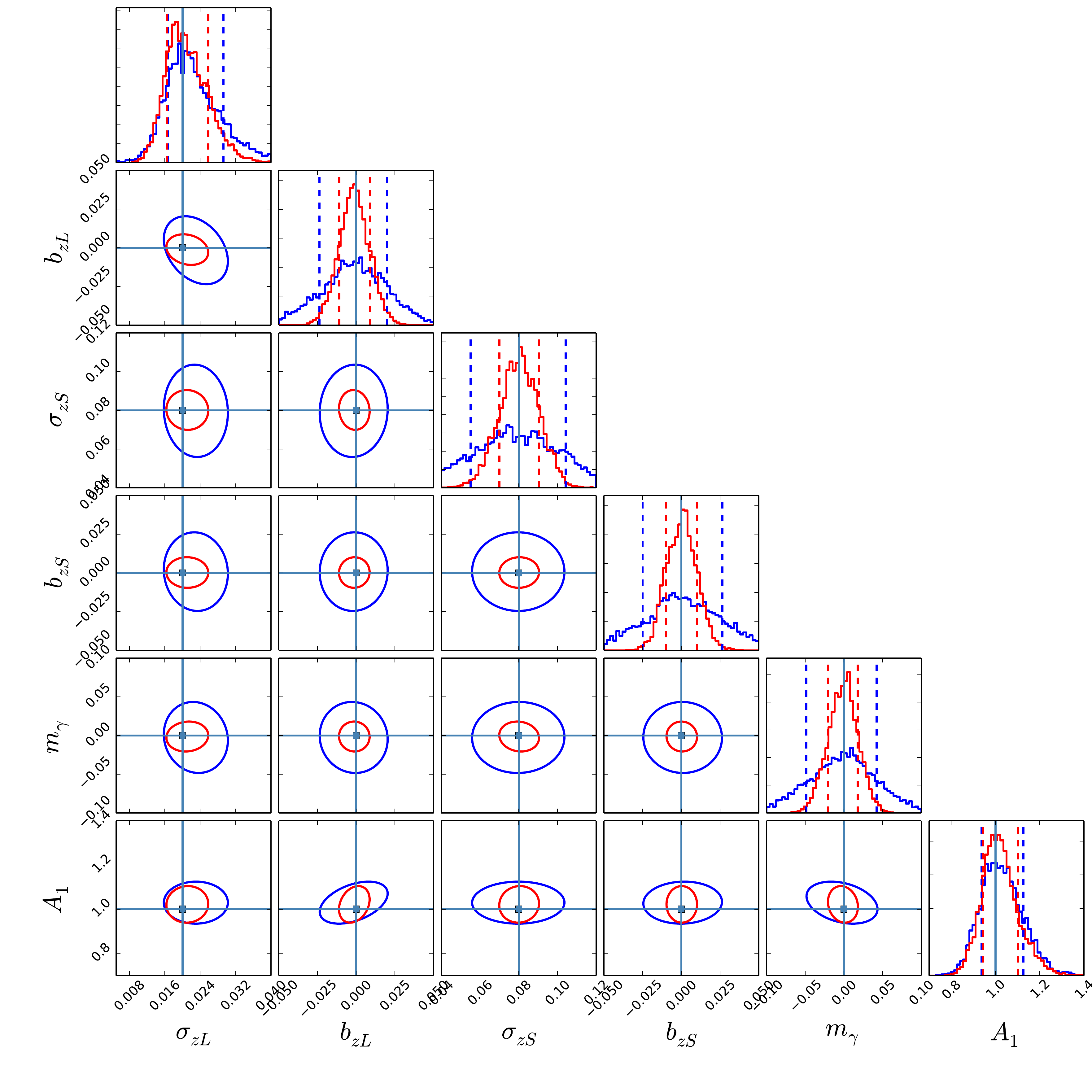}
\caption{Forecast constraints on the systematic effects and growth scaling parameters from the default conservative (red) and relaxed systematic effects priors (blue) Y1 analyses. The relaxed systematic effects priors are a factor of 2.5 weaker than the default priors.}
\label{fig:y1comp} 
\end{figure}

It is also important to understand the effect of systematic effects -- photometric redshift uncertainties and biases for both source and lens galaxies and multiplicative shear calibration -- on the cosmological constraints. This understanding could be used to implement scientific requirements on shear for joint analyses of these types (they may be looser than those needed for cosmic shear) and to estimate the number of spectroscopic redshifts needed to reduce photometric redshift errors. To gauge the impact of systematic effects priors on our final constraining power, we carry out another conservative Y1 analysis with relaxed systematic effects priors, similar to the relaxed HOD priors case above, where we widen the widths of systematic effects priors by a factor of 2.5 compared to the default conservative widths. In Figure~\ref{fig:y1comp}, we compare the default conservative Y1 results, as presented in Figure \ref{fig:y1_consopt_sys}, against the Y1 results with relaxed systematic effects priors. Perhaps the main takeaway is the bottom right panel, which shows that constraints on $A_1$ are degraded minimally (from 7.9\% to 9.4\%) in this case, when the prior constraints on systematic effects parameters are significantly relaxed. This modest degradation is due partly to lack of degeneracy between $A_1$ and most of the systematic effects parameters; note for example the flattened ellipses in the bottom row in columns for $\sigma_{zS}$ and $b_{zL}$. Even though these nuisance parameters are not well-constrained by the data (blue ellipses are much wider than red), they are not degenerate with $A_1$, so they have a limited effect on the final growth constraints. By contrast, the data does constrain $\sigma_{zL}$, the scatter in the lens photometric estimates, quite well even without an external prior. 

\begin{figure}[htpb]
\includegraphics[width=0.6\textwidth]{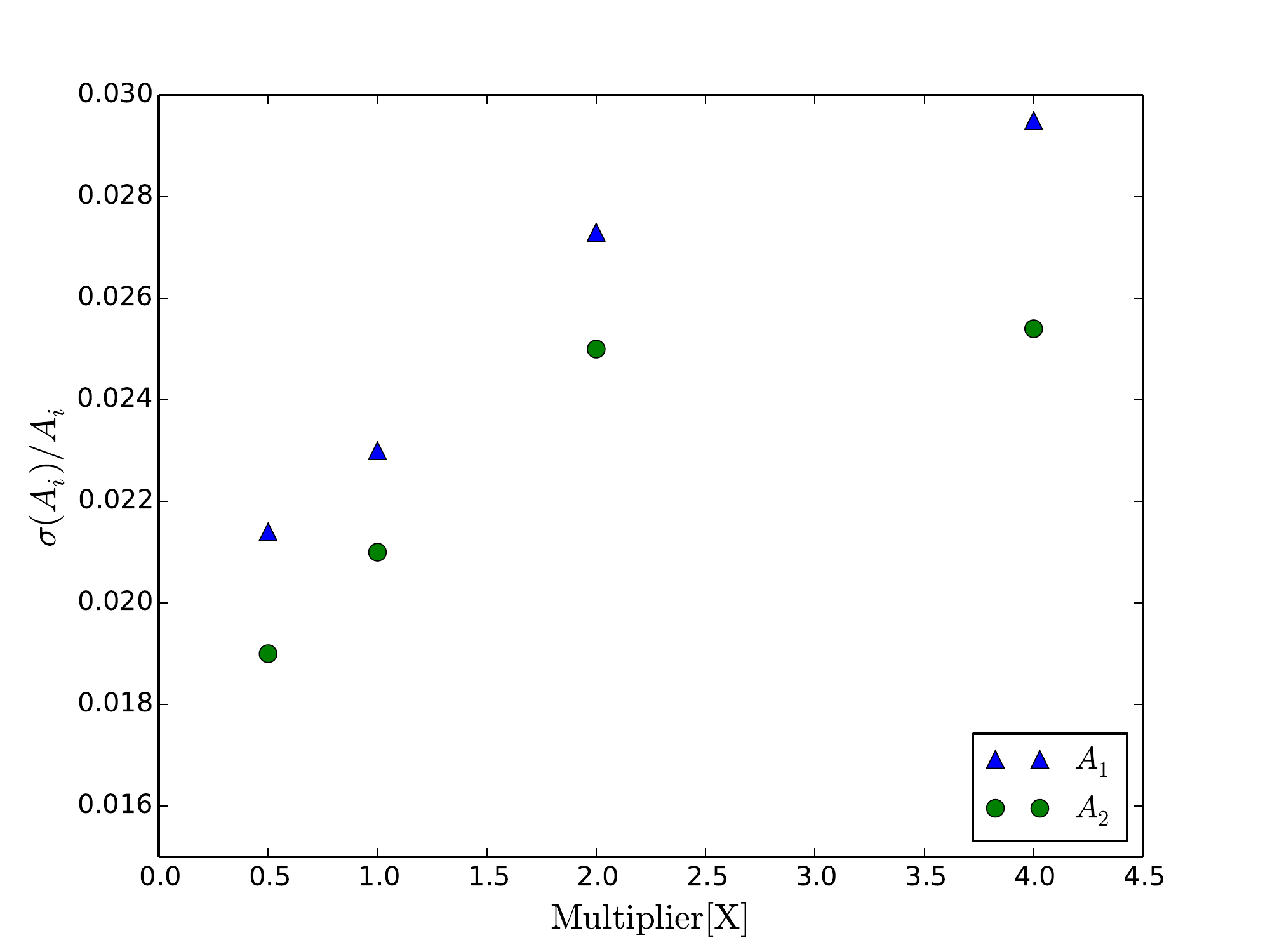}
\caption{Fractional 1-$\sigma$ uncertainty on $A_i$ for different simulated Y5 analyses, plotted against systematic effects prior widths as a factor of the default widths.}
\label{fig:syseffect}
\end{figure}

While the systematic effects parameters had only a small effect in the eventual growth constraints for the conservative Y1 analysis, we anticipate that its relative impact will be bigger in the Y5 scenario, especially for the optimistic Y5 analysis where we employ the tightest HOD priors. To test the significance of systematic effects in the Y5 case, we perform an exercise similar to that for the HOD priors and compare our optimistic Y5 constraints on the growth function with varying prior widths for systematic effects parameters. In Fig. \ref{fig:syseffect}, we present the fractional uncertainty on the growth scaling parameters $A_{1,2}$ with systematic effects priors 0.5, 1, 2, and 4 times the default width. We observe that even under the optimistic Y5 scenario the degradation from relaxing systematic effects priors to growth constraints is weak. As we relax the systematic effects priors, we also found that the lens photo-z parameters ($\sigma_{zL}$ and $b_{zL}$) and the shear calibration parameter $m_\gamma$ constraints do not weaken as much as the relaxation in priors, while the source photo-z parameters ($\sigma_{zS}$ and $b_{zS}$) exhibit changes in constraints that closely follow the relaxed priors. In terms of impact of priors, this indicates that the pipeline is constraining the lens photo-z and the shear calibration parameters by itself, while it is relying on priors to constrain the source photo-z parameters. In terms of growth constraints, this indicates that those same parameters that we are constraining without heavily relying on priors, i.e. lens photo-z and shear calibration parameters, are dominant over those that we constrain largely by priors, i.e. source photo-z parameters, in their impact on our final constraints on the growth function. Finally, the decreasing trend that we observe in Fig. \ref{fig:syseffect} as we tighten our systematic effects priors beyond the default width implies that better understanding and constraining of systematic effects will be important in achieving the tightest possible growth constraints from DES.

\begin{turnpage}
\begin{table}[htp]
\centering
\begin{ruledtabular}
\begin{tabular}{c c c c c c c}
 \multirow{2}{*}{Parameter} & Simulated Y1 & Simulated Y1 & Simulated Y1 & Simulated Y1 & Simulated Y5 & Simulated Y5 \\
& (default conservative) & (default optimistic) & (relaxed HOD priors) & (relaxed systematic effects priors) & (default conservative) & (default optimistic)\\
 \hline
 \noalign{\vskip 1mm}
 $\Omega_M$ 		& $0.312 \pm 0.0033$ & $0.312 \pm 0.0032$ & $0.312 \pm 0.0034$ & $0.312 \pm 0.0033$ & $0.312 \pm 0.0034$ & $0.312 \pm 0.0033$ \\
 $h_0$ 				& $0.674 \pm 0.0032$ & $0.674 \pm 0.0030$ & $0.674 \pm 0.0033$ & $0.674 \pm 0.0033$ & $0.674 \pm 0.0032$ & $0.674 \pm 0.0032$\\
 $10^9 A_s$ 		& $2.150 \pm 0.0048$ & $2.150 \pm 0.0048$ & $2.150 \pm 0.0050$ & $2.150 \pm 0.0049$ & $2.150 \pm 0.0050$ & $2.150 \pm 0.0047$ \\
 
  \hline
 \noalign{\vskip 1mm}
 
 $\log M_\mr{min}$	& $12.36 \pm 0.057$  & $12.36 \pm 0.027$ & $12.36 \pm 0.071$  & $12.36 \pm 0.063$  & $12.36 \pm 0.041$ & $12.36 \pm 0.012$   \\
 $\log M_1$		    & $13.69 \pm 0.100$  & $13.69 \pm 0.019$ & $13.79 \pm 0.238$  & $13.71 \pm 0.098$  & $13.70 \pm 0.051$ & $13.69 \pm 0.008$   \\
 $\sigma_{\log M}$ 	& $0.311 \pm 0.129$  & $0.316 \pm 0.051$ & $0.050 \pm 0.142$  & $0.332 \pm 0.136$  & $0.323 \pm 0.102$ & $0.321 \pm 0.025$   \\
 $\alpha$ 			& $1.278 \pm 0.096$  & $1.281 \pm 0.018$ & $1.221 \pm 0.235$  & $1.263 \pm 0.099$  & $1.279 \pm 0.049$ & $1.280 \pm 0.009$   \\
 $\sigma_{zL}$	    & $0.021 \pm 0.0046$ & $0.020 \pm 0.0042$ & $0.021 \pm 0.0048$ & $0.022 \pm 0.0063$ & $0.020 \pm 0.0027$ & $0.021 \pm 0.0025$  \\
 $b_{zL}$ 			& $0.000 \pm 0.0092$ & $0.000 \pm 0.0096$ & $0.000 \pm 0.0092$ & $-0.001 \pm 0.0218$ & $-0.004 \pm 0.0080$ & $-0.004 \pm 0.0077$  \\
 $A_1$ 				& $1.005 \pm 0.079$  & $1.001 \pm 0.047$ & $1.064 \pm 0.161$  & $1.023 \pm 0.094$  & $1.005 \pm 0.039$ & $1.000 \pm 0.023$  \\
 
 \hline
 \noalign{\vskip 1mm}
 
 $\log M_\mr{min}$	& $12.33 \pm 0.060$  & $12.33 \pm 0.028$ & $12.33 \pm 0.068$  & $12.33 \pm 0.057$  & $12.32 \pm 0.044$ & $12.33 \pm 0.012$   \\
 $\log M_1$		    & $13.59 \pm 0.094$  & $13.58 \pm 0.018$ & $13.68 \pm 0.234$  & $13.59 \pm 0.098$  & $13.58 \pm 0.050$ & $13.58 \pm 0.009$   \\
 $\sigma_{\log M}$ 	& $0.308 \pm 0.136$  & $0.300 \pm 0.049$ & $0.311 \pm 0.140$  & $0.303 \pm 0.135$  & $0.305 \pm 0.122$ & $0.301 \pm 0.023$   \\
 $\alpha$ 			& $1.353 \pm 0.098$  & $1.369 \pm 0.018$ & $1.299 \pm 0.243$  & $1.365 \pm 0.098$  & $1.370 \pm 0.050$ & $1.370 \pm 0.008$   \\
 $\sigma_{zL}$	    & $0.022 \pm 0.0053$ & $0.022 \pm 0.0053$ & $0.021 \pm 0.0057$ & $0.024 \pm 0.0077$ & $0.021 \pm 0.0027$ & $0.022 \pm 0.0030$  \\
 $b_{zL}$ 			& $0.000 \pm 0.0095$ & $0.000 \pm 0.0096$ & $0.001 \pm 0.0096$ & $-0.002 \pm 0.0207$ & $-0.004 \pm 0.0078$ & $-0.003 \pm 0.0078$  \\
 $A_2$ 				& $1.009 \pm 0.078$  & $0.994 \pm 0.048$ & $1.070 \pm 0.162$  & $1.012 \pm 0.084$  & $0.999 \pm 0.039$ & $0.996 \pm 0.021$  \\
 
  \hline
 \noalign{\vskip 1mm}
 
 $\sigma_{zS}$ 		& $0.080 \pm 0.010$  & $0.080 \pm 0.011$ & $0.080 \pm 0.010$  & $0.079 \pm 0.025$  & $0.080 \pm 0.010$ & $0.079 \pm 0.010$   \\
 $b_{zS}$ 			& $0.000 \pm 0.010$  & $0.000 \pm 0.010$ & $0.000 \pm 0.010$  & $0.001 \pm 0.026$  & $0.000 \pm 0.010$ & $0.000 \pm 0.010$   \\
 $m_\gamma$ 		& $0.000 \pm 0.019$  & $-0.001 \pm 0.020$ & $-0.001 \pm 0.020$  & $-0.001 \pm 0.045$  & $-0.001 \pm 0.018$ & $-0.006 \pm 0.019$ \\ 
 
\end{tabular}
\end{ruledtabular}
\caption{Marginalized 1-$\sigma$ bounds on cosmology, HOD, systematic effects, and growth scaling parameters from various simulated Y1 and Y5 analyses. All mass values are in units of $M_\odot / h$.}
\label{tab:bounds}
\end{table}
\end{turnpage}

\clearpage

\section{Discussion}

\begin{figure}[ht]
\includegraphics[width=0.6\textwidth]{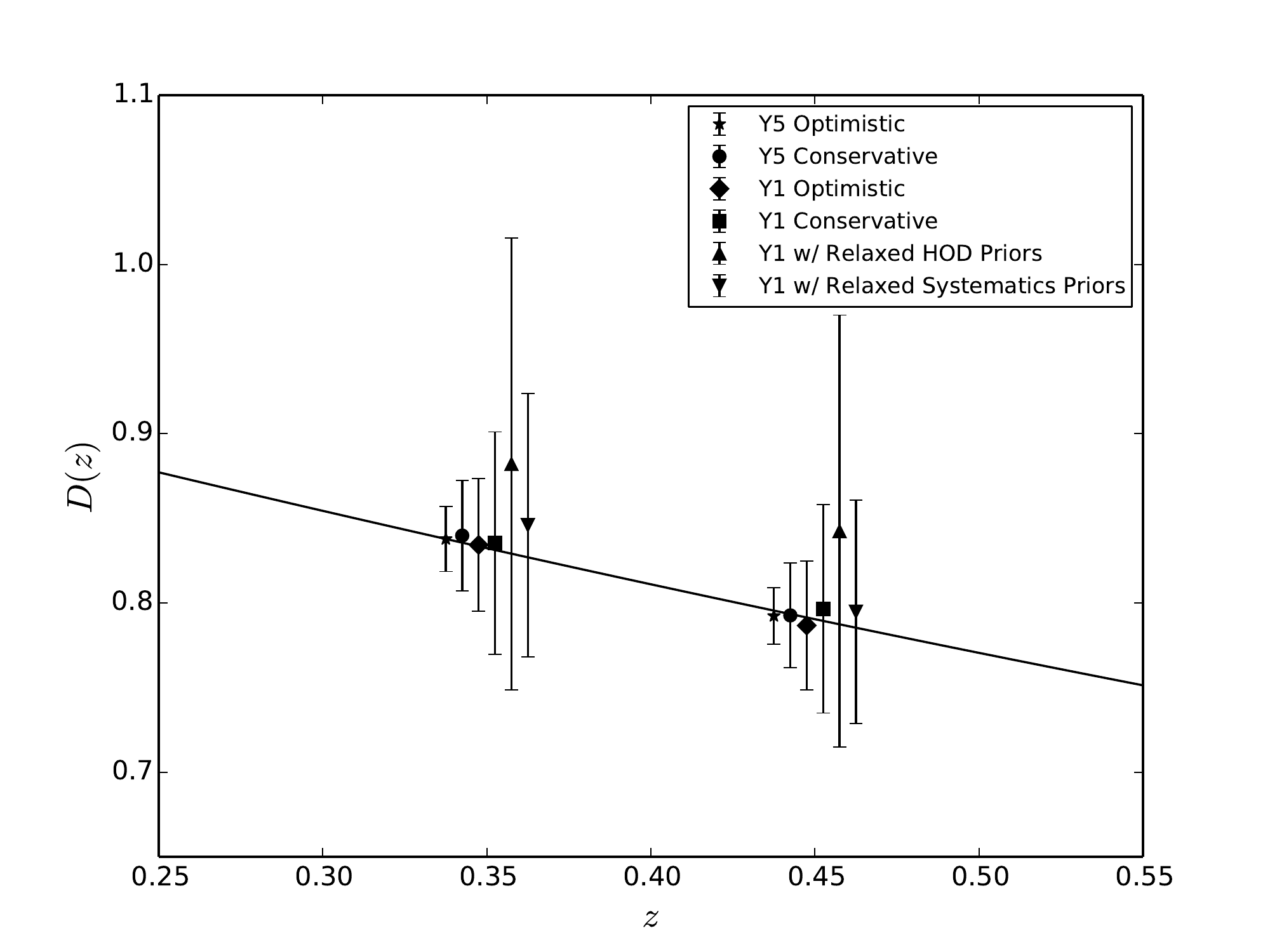}
\caption{Forecasts for the DES 1-$\sigma$ bounds on the growth function $D(z)$ in two different redshift bins, presented for different assumptions on data stage and parameter priors. Points are offset for visibility.}
\label{fig:growth}
\end{figure}

In this paper we demonstrate an implementation of the joint-analysis pipeline for combining galaxy clustering and galaxy-galaxy lensing measurements from photometric surveys. In preparation for DES data analyses, our modeling includes the expected key systematic effects of photometric redshift estimates, shear calibration, and the galaxy-luminosity mass relationship, covering a 20-dimensional parameter space. We show that a joint analysis of large-scale $w(\theta)$ and small-scale $\gamma_t(\theta)$ can conservatively/optimistically constrain the growth function $D(z)$ to within 7.9\%/4.8\% with DES Y1 data and to within 3.9\%/2.3\% with DES Y5 data across two different redshift bins of $0.3<z<0.4$ and $0.4<z<0.5$. These forecasts can be put in the context of existing constraints on the growth function using the abundance of galaxy clusters, weak lensing shear correlations, and redshift space distortions in galaxy clustering. Some recent results include:
\begin{itemize}
\item Galaxy Clusters~\citep{Mantz14}: $\sigma_\mr{8} = 0.83 \pm 0.04$

\item Weak Lensing~\citep{Heymans15}: $\sigma_\mr{8}(\Omega_\mr{m}/0.27)^\alpha = 0.774^{+0.032}_{-0.041}$, $\alpha = 0.46 \pm 0.02$

\item Weak Lensing~\citep{MacCrann15}: $\sigma_\mr{8}(\Omega_\mr{m}/0.3)^{0.5} = 0.81 \pm 0.06$

\item Redshift Space Distortions~\citep{6dfgs}: $\sigma_8 = 0.76 \pm 0.11$

\item Redshift Space Distortions~\citep{wigglez}: $f(z)\sigma_8(z) = 0.413 \pm 0.080, 0.390\pm 0.063, 0.437\pm 0.072$ at $z=0.44,0.6,0.73$

\item Redshift Space Distortions~\citep{Beutler14}: $f(z_\mr{eff})\sigma_8(z_\mr{eff}) = 0.419\pm 0.044$ at $z_\mr{eff} = 0.57$
\end{itemize}
Overall, current constraints are at the 10\% level, implying that DES will soon produce cutting-edge constraints on the growth of structure in the universe with tomography reaching far back in cosmic time, even under conservative assumptions. Fig. \ref{fig:growth} shows the bounds on the $\mr{\Lambda CDM}$ growth function obtained for the two lens bins under the DES Y1 and Y5 specifications considered in this analysis.

An important conclusion is that the HOD parameters are degenerate with our parameters of interest, i.e. the growth scaling parameters $A_i$, but the systematic effects parameters are at most weakly degenerate with $A_i$. By comparing results drawn under different prior settings, we conclude that the final constraining power on the growth function will be driven by our ability to constrain HOD parameters, especially the satellite HOD parameters, as these parameters are both strongly degenerate with $A_i$ and relatively unconstrained without priors. On the other hand, we observe that the central HOD parameters are well constrained without contribution from priors, and also that the systematic effects parameters are either well-constrained or only weakly affecting the final constraining power on $A_i$. The default results are thus strongly driven by the priors on the satellite HOD parameters, but as these prior constraints encompass both conservative and optimistic estimates of the eventual constraining power coming from DES, we believe looking at both the conservative and optimistic forecasts yields reasonable estimates for the ultimate results we can expect from analyses of DES Y1 and Y5 datasets.

\begin{figure}[ht]
\includegraphics[width=0.5\textwidth]{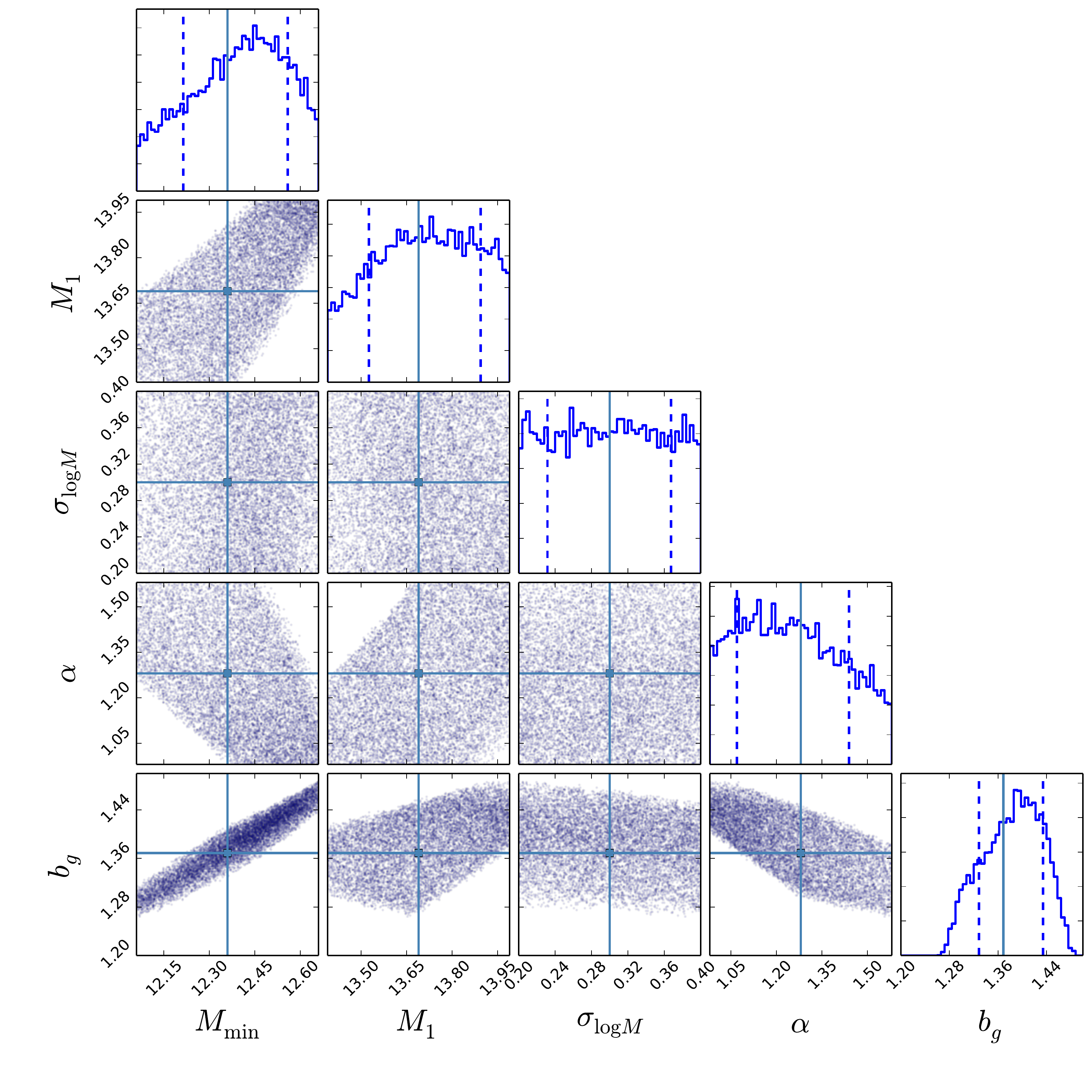}
\caption{Plot of 10,000 random HOD configurations with mean halo masses within 1\% of the fiducial default value for lens bin 1, presented with the galaxy biases derived from these configurations. The plot ranges are set to match the range of parameter values used to generate random HOD configurations. Light blue squares and lines mark the fiducial default HOD for lens bin 1, as presented in Table \ref{tab:params}.}
\label{fig:yscounter}
\end{figure}

The HOD degeneracies also illustrate limitations in the two-step analysis proposed in \citet{Yoo:2012} of (1) determining the (largely cosmology-independent) mean halo mass of the galaxy sample from stacked small-scale lensing measurements and (2) analyzing large-scale galaxy clustering using galaxy bias inferred from the obtained mean halo mass to determine the amplitude of the underlying matter clustering. In this method, the connection from the first to the second step, and consequently the determination of galaxy bias, hinges on a single representative value -- the mean halo mass obtained from galaxy-galaxy lensing. Thus, if galaxy samples with similar mean halo masses can exhibit varying galaxy biases, the two-step approach becomes sub-optimal. And the HOD degeneracies suggest that such a situation is entirely possible. In Fig. \ref{fig:yscounter}, we show 10,000 random HOD configurations with mean halo masses within 1\% of our fiducial default HOD for lens bin 1, along with the derived galaxy bias from those configurations. Note that these are not results from an MCMC analysis, but simply random HOD's within a relatively narrow range of parameter values that yield the desired mean halo masses. Even with this extremely tight requirement in mean halo mass, different galaxy samples exhibit a much wider scatter (up to 10\%) in galaxy bias, as shown in the panel in bottom right corner. This result suggests that the approach employed in our analysis, i.e. a consistent HOD modeling of a given galaxy sample that propagates to predictions for both galaxy clustering and galaxy-galaxy lensing, is a more optimal form of combined probed analysis.

\iffalse
This analysis also shows that a practical implementation of the combined galaxy clustering and galaxy-galaxy lensing analysis method proposed in \citet{Yoo:2012} requires several modifications to the original two-step process of (1) determining the (largely cosmology-independent) halo mass from stacked small-scale lensing measurements, and (2) analyzing large-scale galaxy clustering using the inferred bias to determine the clustering amplitude of matter.  Although the method works, constraints can be improved by modeling the physics on small scales. In practice, optical selection of the lens galaxy sample will produce a broad distribution of halo masses which needs to be accounted for in the modeling of large-scale bias. HOD models are commonly used to characterize this mass distribution,  and their application to galaxy-galaxy lensing and clustering measurements has been demonstrated in many previous studies. Although this modeling is significantly more involved than the more agnostic approach, we have shown that constraints on these parameters can lift degeneracies and strengthen the ultimate constraints on matter clustering. 
\fi

Based on the lessons learned from this study, our current implementation will undergo a number of key improvements in the near future. The most salient improvement will be incorporating small-scale galaxy clustering information. Small-scale galaxy clustering is highly sensitive to satellite galaxies, and thus will allow for tight constraints on the satellite HOD parameters. 
With this improvement, all of our HOD constraints will be data-driven, and our analysis will be self-sufficient without relying on HOD priors. In addition, we expect further validation of our assumed HOD model from a separate DES analysis on HOD modeling. This improved pipeline is anticipated to analyze the DES SVA1 and Y1 data and produce interesting constraints in the near future.

\begin{acknowledgments}

Funding for the DES Projects has been provided by the U.S. Department of Energy, the U.S. National Science Foundation, the Ministry of Science and Education of Spain, 
the Science and Technology Facilities Council of the United Kingdom, the Higher Education Funding Council for England, the National Center for Supercomputing 
Applications at the University of Illinois at Urbana-Champaign, the Kavli Institute of Cosmological Physics at the University of Chicago, 
the Center for Cosmology and Astro-Particle Physics at the Ohio State University,
the Mitchell Institute for Fundamental Physics and Astronomy at Texas A\&M University, Financiadora de Estudos e Projetos, 
Funda{\c c}{\~a}o Carlos Chagas Filho de Amparo {\`a} Pesquisa do Estado do Rio de Janeiro, Conselho Nacional de Desenvolvimento Cient{\'i}fico e Tecnol{\'o}gico and 
the Minist{\'e}rio da Ci{\^e}ncia, Tecnologia e Inova{\c c}{\~a}o, the Deutsche Forschungsgemeinschaft and the Collaborating Institutions in the Dark Energy Survey. 
The DES data management system is supported by the National Science Foundation under Grant Number AST-1138766.

The Collaborating Institutions are Argonne National Laboratory, the University of California at Santa Cruz, the University of Cambridge, Centro de Investigaciones En{\'e}rgeticas, 
Medioambientales y Tecnol{\'o}gicas-Madrid, the University of Chicago, University College London, the DES-Brazil Consortium, the University of Edinburgh, 
the Eidgen{\"o}ssische Technische Hochschule (ETH) Z{\"u}rich, 
Fermi National Accelerator Laboratory, the University of Illinois at Urbana-Champaign, the Institut de Ci{\`e}ncies de l'Espai (IEEC/CSIC), 
the Institut de F{\'i}sica d'Altes Energies, Lawrence Berkeley National Laboratory, the Ludwig-Maximilians Universit{\"a}t M{\"u}nchen and the associated Excellence Cluster Universe, 
the University of Michigan, the National Optical Astronomy Observatory, the University of Nottingham, The Ohio State University, the University of Pennsylvania, the University of Portsmouth, 
SLAC National Accelerator Laboratory, Stanford University, the University of Sussex, and Texas A\&M University.

The DES participants from Spanish institutions are partially supported by MINECO under grants AYA2012-39559, ESP2013-48274, FPA2013-47986, and Centro de Excelencia Severo Ochoa SEV-2012-0234.
Research leading to these results has received funding from the European Research Council under the European Union’s Seventh Framework Programme (FP7/2007-2013) including ERC grant agreements 
 240672, 291329, and 306478.
 
This paper has gone through internal review by the DES collaboration. The DES publication number for this article is DES-2015-0056. The Fermilab preprint number is FERMILAB-PUB-15-311-A.

This work was partially
supported by the Kavli Institute for Cosmological
Physics at the University of Chicago through grants NSF
PHY-1125897 and an endowment from the Kavli Foundation
and its founder Fred Kavli. The work of SD is
supported by the U.S. Department of Energy, including
grant DE-FG02-95ER40896.

\end{acknowledgments}

\clearpage

\bibliography{references}

\newpage

\appendix
\section{Forecast Figures}
\label{sec:figs}

\begin{figure}[ht]
\includegraphics[width=\textwidth]{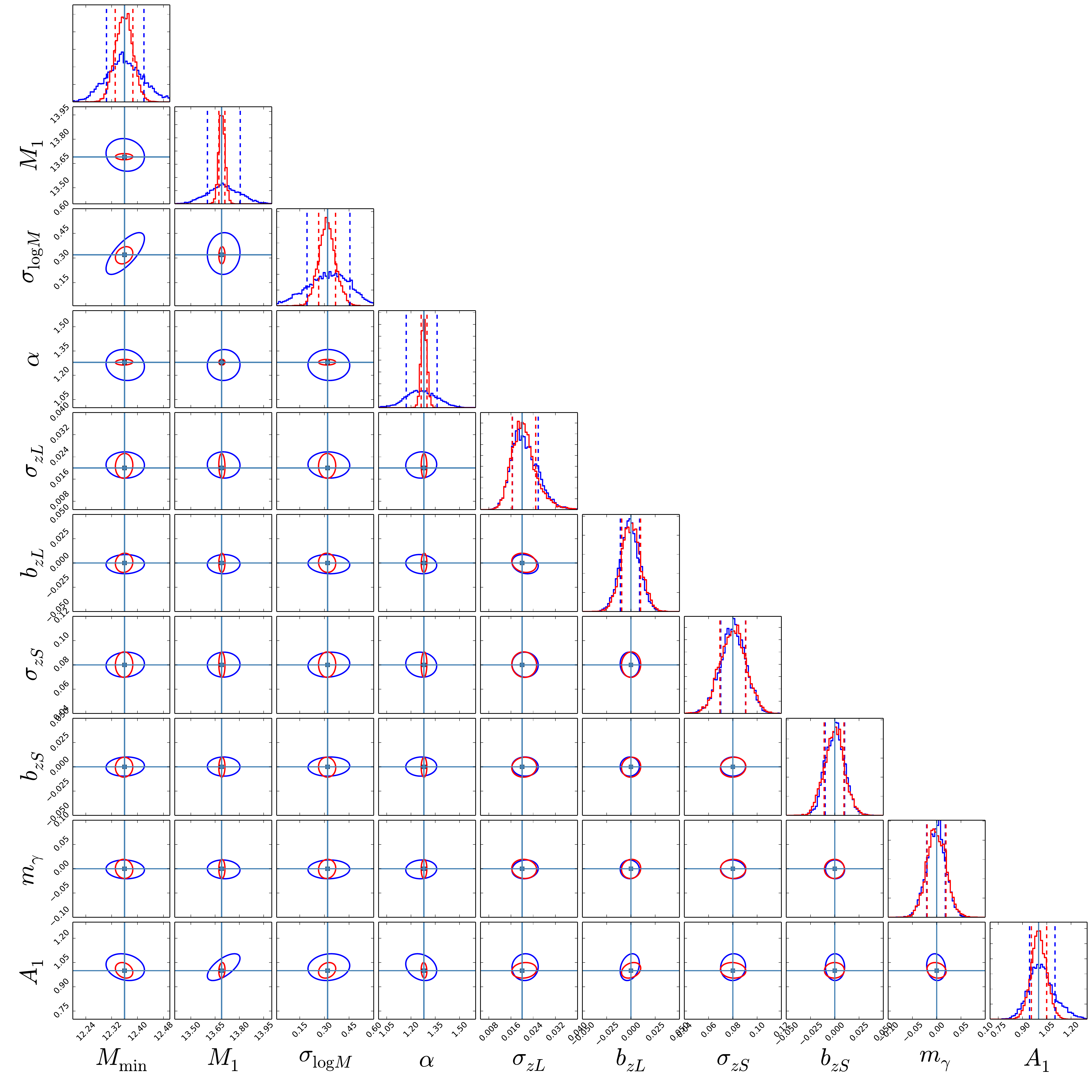}
\caption{Marginalized 1D parameter constraints with dotted vertical lines at $\pm 1\sigma$ (diagonal) and 2D 1-$\sigma$ confidence ellipses (off-diagonal), representing the conservative (blue) and optimistic (red) Y1 parameter constraint forecasts for the first lens bin. Vertical and horizontal axes represent the 4 HOD parameters, 5 systematic effects parameters, and the growth scaling parameter, respectively. Light blue lines and squares correspond to the true values used in generating the simulated measurement vector.}
\label{fig:contours}
\end{figure}

\begin{figure}[ht]
\includegraphics[width=\textwidth]{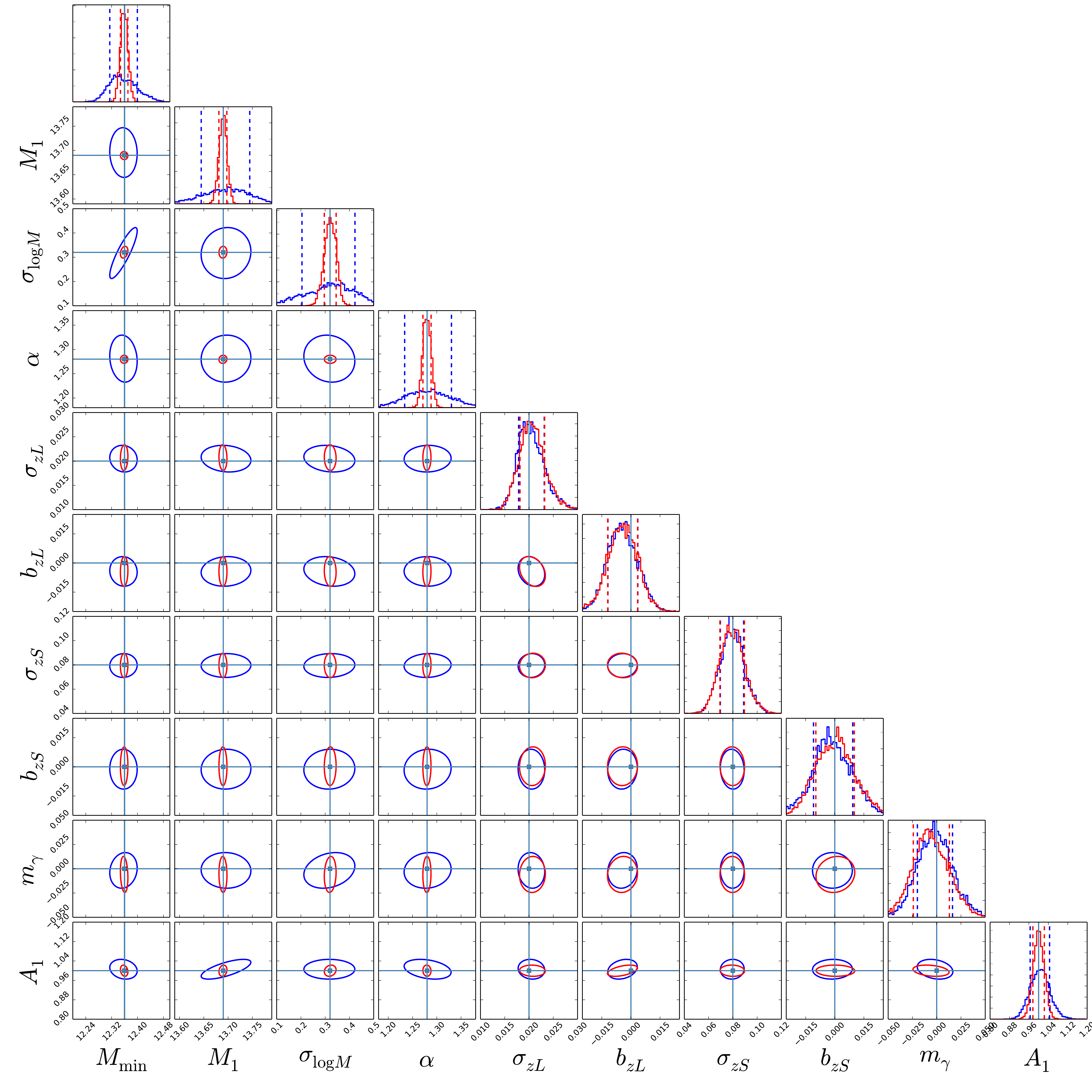}
\caption{Marginalized 1D parameter constraints with dotted vertical lines at $\pm 1\sigma$ (diagonal) and 2D 1-$\sigma$ confidence ellipses (off-diagonal), representing the conservative (blue) and optimistic (red) Y5 parameter constraint forecasts for the first lens bin. Vertical and horizontal axes represent the 4 HOD parameters, 5 systematic effects parameters, and the growth scaling parameter, respectively. Light blue lines and squares correspond to the true values used in generating the simulated measurement vector.}
\label{fig:contours2}
\end{figure}

\end{document}